\documentclass{ar-1col-S2O}
\usepackage[numbers]{natbib}
\usepackage{url}
\usepackage{anyfontsize}
\usepackage{graphicx}
\usepackage{epstopdf}
\usepackage{subfig}
\usepackage{epsfig}
\usepackage{amsmath,amssymb,amsfonts}
\usepackage{color}
\usepackage{xcolor}
\usepackage{url}
\usepackage[utf8]{inputenc}
\usepackage[bookmarks,
                bookmarksopen = true,
                bookmarksnumbered = true,
                linktocpage,
                colorlinks = true,
                linkcolor = blue,
                urlcolor  = blue,
                citecolor = blue,
                anchorcolor = green,
                hyperindex = true,
                hyperfigures]
                {hyperref}
\allowdisplaybreaks

\begin{document}

\markboth{Cao and Qin}{Medium response}

\title{Medium Response and Jet-Hadron Correlations in Relativistic Heavy-Ion Collisions}

\author{Shanshan Cao$^1$ and Guang-You Qin$^2$
\affil{$^1$Institute of Frontier and Interdisciplinary Science, Shandong University, Qingdao, Shandong 266237, China; email: shanshan.cao@sdu.edu.cn}
\affil{$^2$Institute of Particle Physics and Key Laboratory of Quark and Lepton Physics (MOE), Central China Normal University, Wuhan, Hubei 430079, China; email: guangyou.qin@ccnu.edu.cn}}

\begin{abstract}
High-energy heavy-ion collisions at the Relativistic Heavy-Ion Collider and the Large Hadron Collider have evolved from qualitative understanding into precise extraction of the properties of the Quantum Chromodynamics medium at extremely high temperature. Jet quenching has offered unique insights into the transport properties of the Quark-Gluon Plasma (QGP) created in these energetic collisions. Apart from medium modification of jets, jet-induced medium excitation constitutes another crucial aspect of jet-QGP interaction, and is indispensable in understanding the soft components of jets. We review recent theoretical and phenomenological developments on medium response to jet energy loss, including an overview of both weakly and strongly coupled approaches for describing the thermalization and propagation of energy deposition from jets, effects of medium response on jet observables, and exploration of its unique signatures in jet-hadron correlations. Jet-induced medium excitation is shown to be an essential component in probing the in-medium dynamics of jets and a critical step towards precise extraction of the QGP properties.
\end{abstract}


\maketitle

\tableofcontents

\section{INTRODUCTION}
\label{sec:introduction}

A wealth of experimental data from high-energy nucleus-nucleus collisions performed at the Relativistic Heavy-Ion Collider (RHIC) and the Large Hadron Collider (LHC) have revealed that the hot and dense nuclear matter produced in these energetic collisions is a strongly-interacting quark-gluon plasma (QGP). Among various probes of the nuclear matter, jet quenching provides smoking gun evidence of the QGP~\cite{Wang:1991xy, Qin:2015srf, Majumder:2010qh, Blaizot:2015lma, Cao:2020wlm}. A jet is a spray of collimated particles (typically hadrons) that emanate from hard scatterings between nucleons or nuclei. It starts with a highly virtual (off-shell) parton produced in these collisions, and then develops with successive splittings from this parton and hadronization of the daughter partons (jet partons). Before hadronization, a (partonic) jet is composed of a collection of jet partons. In heavy-ion collisions, high-energy jet partons suffer energy loss via elastic and inelastic interactions with the QGP before being observed.
Various theoretical formalisms have been developed for studying collisional energy loss~\cite{Bjorken:1982tu, Braaten:1991we, Djordjevic:2006tw, Qin:2007rn} and medium-induced radiation~\cite{Baier:1994bd, Baier:1996kr, Zakharov:1996fv, Gyulassy:1999zd, Wiedemann:2000za, Arnold:2002ja, Wang:2001ifa} experienced by jet partons propagating through a dense nuclear matter.
A direct consequence of parton energy loss is the suppression of high transverse momentum ($p_\mathrm{T}$) hadron and jet spectra as compared to the expectation from independent proton-proton collisions~\cite{ATLAS:2014ipv, CMS:2016xef, ALICE:2018vuu, Bass:2008rv}.
In addition, parton energy loss can lead to modification of jet-related correlations, such as dijet and $\gamma/Z$-jet asymmetries, dihadron and $\gamma/Z$-hadron correlations~\cite{ATLAS:2010isq, CMS:2012ytf, Zhang:2007ja, Qin:2009bk, Qin:2010mn, Chen:2016vem, Chen:2017zte, Luo:2018pto, Zhang:2018urd}.
Jet-medium interactions can also change the internal structure of full jets, such as jet shape and jet fragmentation function \cite{CMS:2012nro, CMS:2013lhm, ATLAS:2014dtd, STAR:2021kjt, Chang:2016gjp, Tachibana:2017syd, Chang:2019sae, JETSCAPE:2023hqn}.
One of the main goals of jet quenching study is to understand the detailed mechanisms of jet-medium interactions and ultimately extract various properties of the QGP produced in high-energy heavy-ion collisions.

\begin{marginnote}[]
\entry{$\hat{q}$}{the rate of transverse momentum broadening ($d\langle k_\perp^2\rangle/dL$) via elastic scatterings.}
\end{marginnote}

Extensive studies on jet quenching in heavy-ion collisions have been performed in the past decade, which has rendered rich information about the hot and dense QGP~\cite{Majumder:2011uk,Blaizot:2013hx,Djordjevic:2014tka,Xing:2019xae,Huss:2020dwe,Caucal:2018dla,Mehtar-Tani:2021fud,Zhao:2021vmu,Liu:2021izt,JETSCAPE:2022jer,Zhang:2022rby}.
With a vast amount of experimental data available at RHIC and the LHC, jet quenching studies have entered a precision era.
For example, the comprehensive efforts from JET and JETSCAPE Collaborations have discovered that the jet transport coefficient $\hat{q}$ in the QGP is about two orders of magnitude larger than the value in cold nuclei~\cite{JET:2013cls, JETSCAPE:2021ehl}.
Since $\hat{q}$ can be written as the correlation of the gluon fields, its value can reflect the gluon density and other transport properties of the dense nuclear medium~\cite{Baier:1996sk}.
Interestingly, Ref.~\cite{Majumder:2007zh} shows that the jet quenching parameter $\hat{q}/T^3$ is also related to the specific shear viscosity of the QGP, with $T$ being the medium temperature.
Therefore, determining the temperature dependence of $\hat{q}$ can shed light on how the QCD matter evolves from a weakly-interacting gas at sufficiently high temperature into a strongly-coupled fluid at the temperature range achieved at current RHIC and LHC experiments. Recently, exploration of $\hat{q}$ has also been extended to finite chemical potential and shown to be a novel probe of the critical point of the QCD phase diagram~\cite{Wu:2022vbu}.

\begin{marginnote}[]
\entry{Specific viscosity}{also known as the shear-viscosity-to-entropy-density ratio ($\eta_\mathrm{v}/s$).}
\end{marginnote}

Apart from medium modification of jets, how the medium responds to the lost energy from jets is another crucial aspect of jet-medium interactions. The latter is known as jet-induced medium excitation, or medium response.
Since jet partons can travel much faster than the speed of sound of the QGP, one expects a Mach cone like structure of the medium response excited by jet propagation~\cite{CasalderreySolana:2004qm, Stoecker:2004qu, Chaudhuri:2005vc, Ruppert:2005uz, Gubser:2007ga, Chesler:2007an, Neufeld:2008dx}.
Earlier studies have shown that the detailed structure of a jet-induced Mach cone is very sensitive to the equation of state and transport properties, such as the shear viscosity, of the QGP.
Therefore, if the jet-excited Mach cone can be observed in heavy-ion experiments, it will provide a direct probe of these medium properties.
However, there are a lot of complications with jet-medium interactions, which makes the direct detection of jet-induced flow and Mach cone extremely difficult.
For example, the strong radial flow of the hydrodynamically expanding medium can significantly distort the Mach cone structure~\cite{Ma:2010dv, Betz:2010qh, Tachibana:2015qxa}.
In addition, a jet is not a single leading parton, but contains many shower partons as well.
While the shower partons can serve as additional sources of energy deposition into the medium and make stronger medium excitations than a single parton does~\cite{Qin:2009uh,Neufeld:2009ep}, the energy-momentum deposition by shower partons typically has a broad distribution in both momentum and coordinate spaces and also fluctuates from event to event~\cite{Neufeld:2011yh, Renk:2013pua}.
Furthermore, the medium is also event-by-event fluctuating, therefore, one is searching for the medium response signal on top of a large and fluctuating background.

\begin{marginnote}[]
\entry{Parton shower}{successive splittings of energetic partons created in the initial hard scatterings. Partons produced from these splittings are called ``shower partons".}
\end{marginnote}

Despite these difficulties, tremendous efforts have been devoted to studying jet-induced medium response and seeking its unique signals in relativistic heavy-ion collisions.
For example, Ref.~\cite{Tachibana:2017syd} builds a coupled jet-fluid model to study jet propagation and medium response.
It is found that jet-induced medium flow and excitations can diffuse to large angles with respect to the jet direction.
After the inclusion of medium response, the model can naturally explain the enhancement of the fractional energy of a jet within a given annulus, or jet shape, at large radius~\cite{CMS:2016cvr}.
Similar findings have also been obtained in other calculations from the \textsc{Lbt}, \textsc{Jewel} and \textsc{Martini} models~\cite{Luo:2018pto, KunnawalkamElayavalli:2017hxo, Park:2018acg}, in which parton transport approaches are used to approximate the medium response effect.
References~\cite{Chen:2017zte, Chen:2020tbl} have developed a comprehensive \textsc{CoLbt-Hydro} model in which the evolutions of jet partons and the excited QGP medium are simulated concurrently.
By applying the \textsc{CoLbt-Hydro} model to $\gamma$-triggered hadrons and jets, it is found that while parton energy loss usually leads to the reduction of the high $p_\mathrm{T}$ hadron yield, jet-induced medium excitation can lead to the enhancement of low $p_\mathrm{T}$ particles.
Reference~\cite{Chen:2020tbl} predicts the depletion of energy behind the hard jet, known as the ``diffusion wake", can result in a suppression of the soft hadron production on the side opposite to the jet direction.
However, the CMS experiment has observed an enhancement of soft hadrons in both $Z$ and jet directions in $Z$-hadron correlations~\cite{CMS:2021otx}. This can be caused by the multi-parton interactions (MPI) in nuclear collisions. Later, Ref.~\cite{Yang:2021qtl} proposes to use the transverse momentum imbalance between the trigger particle and its associated jet to localize the initial jet positions~\cite{Zhang:2007ja, Zhang:2009rn, He:2020iow}, which can enhance the signal of diffusion wake in $Z/\gamma$-jet events. The machine learning technique has also been introduced~\cite{Yang:2022yfr} to identify jets from specific locations and along particular paths, leading to a more direct search for signals of medium response. Very recently, Ref.~\cite{Luo:2021voy} proposes that the enhanced ratio of baryons to mesons at intermediate $p_\mathrm{T}$ around the quenched jets is a unique signature of jet-induced medium excitation. An increase of jet flavors due to jet-medium scatterings is also found in Ref.~\cite{Sirimanna:2022zje} using various model calculations.

In this article, we review recent developments in the study of jet-induced medium excitation in relativistic heavy-ion collisions. We first provide a short review on how various models implement medium response to jet quenching. Then we discuss how medium response manifests in the final state jet observables, including jet spectra, jet anisotropy and jet structure/substructure. In the end, we present some recent studies on searching for signals of medium response via jet-hadron correlations.

\section{MODEL IMPLEMENTATIONS OF MEDIUM RESPONSE}

\subsection{Hydrodynamic response to energy-momentum deposition}
\label{subsec:hydroResponse}

In ultra-relativistic heavy-ion collisions, baryons from projectile and target nuclei rapidly penetrate each other, leaving a baryonic free QGP at mid-rapidity. In the absence of net baryon number, the energy-momentum conservation requires
\begin{equation}
\label{eq:hydroEq}
\partial_\mu T^{\mu\nu} = 0,
\end{equation}
where $T^{\mu\nu}=\varepsilon u^{\mu}u^{\nu}-(P+\Pi) \Delta^{\mu\nu} + \pi^{\mu\nu}$ is the energy-momentum tensor, with $\varepsilon$, $u^{\mu}$, $P$, $\Pi$ and $\pi^{\mu\nu}$ being the local energy density, fluid four-velocity, pressure, bulk pressure and shear stress tensor respectively, and $\Delta^{\mu\nu}=g^{\mu\nu}-u^{\mu}u^{\nu}$ being the projection operator orthogonal to the fluid velocity. By convention, the metric tensor is taken as $g^{\mu\nu}=(1,-1,-1,-1)$ in the Cartesian coordinates,
while the Milne coordinates are widely used in relativistic nuclear collisions, with  $g^{\mu\nu}=(1,-1,-1,-\tau^2)$. Based on Eq.~(\ref{eq:hydroEq}), hydrodynamic models~\cite{Romatschke:2009im,Heinz:2013th,Shen:2020mgh} have been developed to describe the spacetime evolution profile of the QGP. In additional to studying the bulk properties of the hot nuclear matter, these hydrodynamic models also provide essential information for investigating jet-medium interactions, such as the temperature and flow velocity of a fluid cell through which a jet parton plows at a given time.

\begin{marginnote}[]
\entry{Milne coordinate}{$(\tau,x,y,\eta_\mathrm{s})$ with $\tau=\sqrt{t^2-z^2}$ and $\eta_\mathrm{s}=(1/2)\ln[(t+z)/(t-z)]$}, where $(t,x,y,z)$ are the Cartesian coordinates.
\end{marginnote}

Jet-medium interactions include both medium modification of jets and jet-induced medium excitation. The latter is the focus of this review. With the assumption that the lost energy from jet partons have reached local thermalization with the surrounding QGP, further evolution of the medium with this energy-momentum deposition can be described by a hydrodynamic equation with a source term $J^\nu$,
\begin{equation}
\label{eq:hydroSource}
\partial_\mu T^{\mu\nu}(x)=J^\nu(x),
\end{equation}
where $J^\nu(x)=[dE/d^4x,d{\vec p}/d^4x]$ represents the space-time distribution of energy ($E$) and momentum ($\vec{p}$) transferred from jets to medium. If we assume the energy-momentum deposition is a small perturbation to the QGP, Eq.~(\ref{eq:hydroSource}) can be linearized as~\cite{CasalderreySolana:2004qm,Neufeld:2008fi,Neufeld:2008dx}
\begin{equation}
\label{eq:linearHydro}
T^{\mu\nu}\approx T^{\mu\nu}_0+\delta T^{\mu\nu};\;\; \partial_\mu T^{\mu\nu}_0=0, \;\; \partial_\mu \delta T^{\mu\nu}=J^\nu;
\end{equation}
where $T^{\mu\nu}_0$ represents the unperturbed energy-momentum tensor that still follows Eq.~(\ref{eq:hydroEq}), while the variation part $\delta T^{\mu\nu}$ describes the medium response to jet energy loss.

To obtain the evolution of this perturbation in spacetime, one may first decompose it as $\delta T^{00}\equiv\delta \epsilon$, $\delta T^{0i}\equiv g^i$, and $\delta T^{ij}=\delta^{ij}c_s^2\delta\epsilon + \frac{3}{4}\Gamma_s (\partial^i g^j + \partial^j g^i + \frac{2}{3} \delta^{ij}  \nabla\cdot{\vec g})$, where $\delta\epsilon$ is the excess energy density compared to the unperturbed medium, ${\vec g}$ is the momentum current, $c_s$ is the speed of sound, $\Gamma_s\equiv 4\eta_\mathrm{v}/[3(\epsilon_0+p_0)]$ is the sound attenuation length with $\epsilon_0$ and $p_0$ being the unperturbed local energy density and pressure respectively. Then, the last part of Eq.~(\ref{eq:linearHydro}) can be solved algebraically with Fourier transformation into the momentum space $(\omega, \vec{k})$. For instance, the excess energy density reads
\begin{equation}
\label{eq:linearHydroSoln1}
\delta\epsilon(\omega,{\vec k}) = \frac{(i\omega-\Gamma_s k^2)J^0(\omega,{\vec k})+ikJ_\mathrm{L}(\omega,{\vec k})}{\omega^2-c_s^2 k^2 + i\Gamma_s\omega k^2},
\end{equation}
in which the source term is decomposed into transverse and longitudinal components as ${\vec J}={\vec J}_\mathrm{T} + \hat{k}{J}_\mathrm{L}$. The momentum current $\vec{g}$ can be achieved in the same manner. In the end, these solutions are Fourier transformed back to the coordinate space.

\begin{figure}[tbp]
    \centering
    \includegraphics[width=0.9\textwidth]{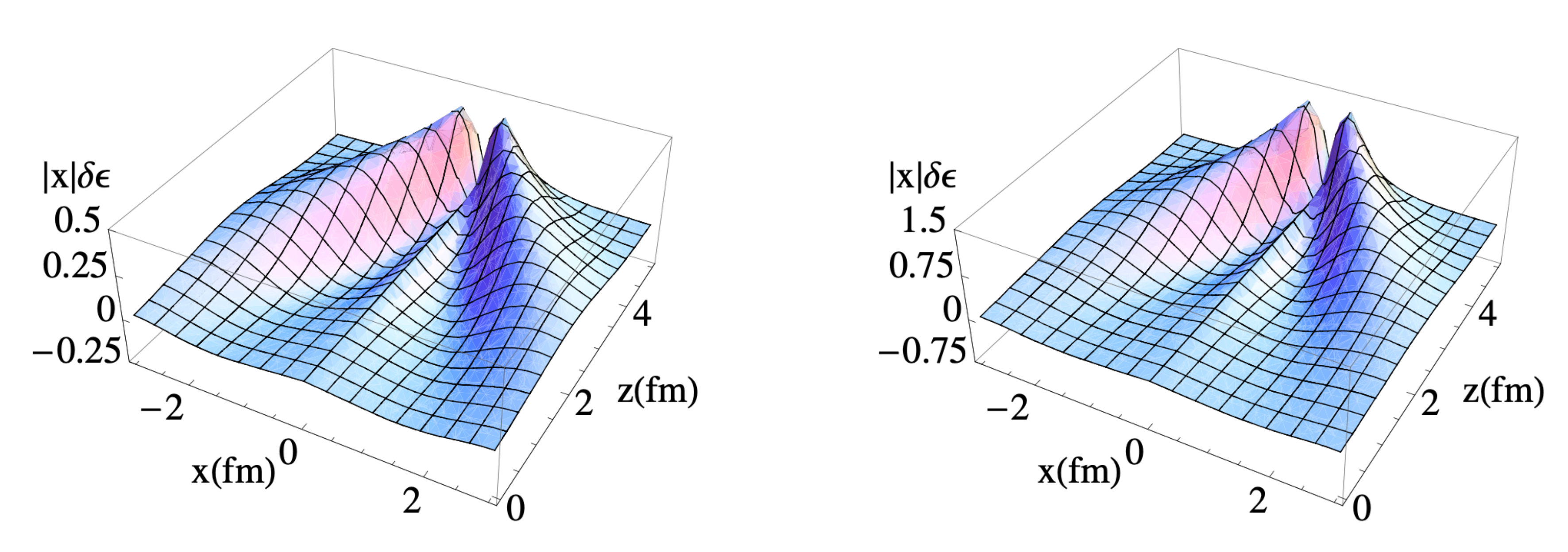}
    \caption{Hydrodynamical response to energy deposition from a single quark (left) {\it vs.} a quark-initiated jet shower (right). The figures are from Ref.~\cite{Qin:2009uh}.}
    \label{fig:Qin-shower-source}
\end{figure}

Based on this linearized hydrodynamics approach, the spacetime structure of the jet-induced medium excitation has been investigated in many earlier studies. Since high-energy partons travel with supersonic speed, it has been proposed in Ref.~\cite{CasalderreySolana:2004qm} that the lost energy of quenched jets appears in the form of collective hydrodynamic motion similar to ``sonic booms", leading to a Mach cone pattern of emitted particles around the quenched jets. As illustrated in Fig.~\ref{fig:Qin-shower-source}, when a jet propagates along the $\hat{z}$-direction, it induces a conic distribution of enhanced energy around its path. Meanwhile, depletion of energy, known as the ``diffusion wake", can be observed in the backward direction. This pattern has been further investigated in Refs.~\cite{Neufeld:2008fi,Neufeld:2008dx}, and is found to become weaker as the specific viscosity of the medium increases. An increasing viscosity can reduce the jet energy loss rate and damp the sound propagation inside the medium, both suppressing the intensity of Mach cone. Within a mode-by-mode formalism of solving the hydrodynamic response~\cite{Yan:2017rku}, the former effect has been found to be more dominant compared to the latter. Different structures of jet-induced medium excitations inside weakly and strongly coupled QGP are also compared in Ref.~\cite{Ruppert:2005uz}. Therefore, if measured, the Mach cone structures may provide direct constraints on the transport parameters of the QGP, including its shear viscosity and the speed of sound.

While most earlier studies constructed the source term using the energy-momentum deposition from a single parton, extension to that from a full parton shower has been proposed in Refs.~\cite{Qin:2009uh, Neufeld:2009ep}, in which each parton within jets is allowed to transfer energy into the medium through a combination of elastic and inelastic processes. As shown in Fig.~\ref{fig:Qin-shower-source}, this more realistic modeling of the source term (right panel) leads to a significant enhancement of the Mach cone structure (see the vertical scales) as compared to earlier studies based on energy loss from a single parton (left panel). On the other hand, quantum interference between the primary jet parton and its emitted gluons is found to suppress the multi-parton source term in the forward direction of jet propagation while causing enhancement at large angles, and in the end may spoil the conic structure of medium response~\cite{Neufeld:2011yh}.

The linear response approach [Eq.~(\ref{eq:linearHydro})] treats the energy deposition from hard partons as a perturbation to the QGP, which may become invalid for high-energy jets and low-temperature medium. Full solutions to  hydrodynamic equations with source terms [Eq.~(\ref{eq:hydroSource})] have been developed using (1+1)-D~\cite{Floerchinger:2014yqa}, (2+1)-D~\cite{Chaudhuri:2005vc} and (3+1)-D~\cite{Betz:2010qh,Tachibana:2014lja,Tachibana:2020mtb} hydrodynamic models. Here, in order to isolate jet effects on the medium, one needs to subtract the solution of hydrodynamic equations without the source term from the full simulation.

As we move to lower energy heavy-ion collisions, effects of finite baryon chemical potential become important. Recently, jet-induced medium excitation has been extended to a baryon-charged medium in Ref.~\cite{Du:2022oaw}. In this case, conservation of the baryon charge, $\partial_\mu N^\mu (x)=\rho_\mathrm{B}(x)$,
should be solved together with Eq.~(\ref{eq:hydroSource}), where $\rho_\mathrm{B}$ represents the baryon charge transferred from jets to medium. Within this framework, Mach cone and wake structures are observed for both the energy density and the baryon density disturbed by jet propagation, even with $\rho_\mathrm{B}$ set to zero. These features can be reflected in the angular distributions of particle production in the transverse plane, with sizable differences observed between baryons and mesons, as well as between baryons and anti-baryons.

\subsection{Weakly coupled approach for medium response}
\label{subsec:recoil}

The hydrodynamic description of jet-induced medium excitation assumes thermal equilibrium of the jet energy deposition with its surrounding QGP medium. However, for energy deposition much above the thermal scale, one may treat it as a quasi-particle, whose evolution can be described within the perturbative picture in the same way as for jet partons.

For instance, a linear Boltzmann transport model (\textsc{Lbt})~\cite{Wang:2013cia,Cao:2016gvr,Xing:2019xae} has been developed for studying scatterings of jet partons through a hydrodynamic medium. The phase space distribution of jet partons (denoted as $a$) evolves according to the Boltzmann equation as
\begin{equation}
  \label{eq:boltzmann1}
  p_a\cdot\partial f_a(x_a,p_a)=E_a (\mathcal{C}_a^\mathrm{el}+\mathcal{C}_a^\mathrm{inel}),
\end{equation}
where $x_a=(t_a, \vec{x}_a)$ and $p_a=(E_a,\vec{p}_a)$ are the four-position and four-momentum, $\mathcal{C}_a^\mathrm{el}$ and $\mathcal{C}_a^\mathrm{inel}$ are the collision integrals for elastic and inelastic scatterings respectively. In \textsc{Lbt}, elastic scatterings include all possible $ab\rightarrow cd$ channels at the leading order, where the medium parton $b$ follows the thermal distribution inside the QGP, with its local temperature and flow velocity provided by hydrodynamic calculations described in the previous subsection. Inelastic scatterings in \textsc{Lbt} are associated with the medium-induced gluon bremsstrahlung process, where the spectra of emitted gluons are calculated within the higher-twist energy loss formalism~\cite{Guo:2000nz,Majumder:2009ge}.

Starting from hard partons produced by the initial energetic scatterings and their vacuum showers, \textsc{Lbt} generates a tree of their daughter partons based on their elastic and inelastic scatterings with the medium. The final state ($c$) of an incoming parton ($a$), together with the emitted gluons, constitute ``jet partons" in the \textsc{Lbt} model. Meanwhile, thermal partons are scattered out of the QGP by hard partons, causing a depletion of energy inside the medium. We call the final state ($d$) of these thermal partons ``recoil partons", and the initial state ($b$) ``negative partons". The latter is used to represent the energy depletion inside the QGP. In this approach, recoil and ``negative" partons constitute the jet-induced medium excitation. The recoil partons are allowed to re-scatter with the QGP in the same way as jet partons do, mimicking the further evolution of medium response within the perturbative picture, and their offsprings are also labelled as recoil or ``negative" partons. Note that in order to guarantee the energy-momentum conservation of the entire system, the energy-momentum of ``negative" partons should be subtracted in the end when analyzing jet observables.

\begin{marginnote}[]
\entry{``Negative" partons}{energy-momentum depletion from the QGP by jet scatterings, also named as ``back-reaction" in literature.}
\end{marginnote}

Using this recoil method, one can also find the Mach cone structure of the jet-induced medium excitation. For instance, in Ref.~\cite{He:2015pra}, a shock wave pattern of the parton energy distribution has been observed around the propagation direction of a high-energy jet parton. This shock wave diffuses quickly during its further interactions with the QGP, due to the large value of shear viscosity within a perturbative scattering picture. Besides the enhanced wave front, depletion of the medium energy -- diffusion wake -- can also be seen behind the jet. These features exist only when both recoil and ``negative" partons have been taken into account, and they are in qualitative agreement with results from hydrodynamic calculations of medium response, as previously shown in Fig.~\ref{fig:Qin-shower-source}.

The recoil description of medium response has also been applied in the \textsc{Martini}~\cite{Schenke:2009gb,Park:2018acg} and \textsc{Jewel}~\cite{Zapp:2011ya,Zapp:2012ak,Zapp:2013vla} event generators.
Besides different modelings of inelastic scatterings, detailed implementations of medium response are also slightly different. In \textsc{Martini}, only recoil partons above certain kinematic threshold (e.g. 4$T$) are kept, and are allowed to re-scatter with the QGP. Recoil partons below the threshold, together with ``negative" partons are considered part of the medium background and are not traced in the transport simulation. In contrast, \textsc{Jewel} includes both recoil and ``negative" partons, although secondary scatterings between the recoil partons and the medium have not been taken into account.

A similar method of medium response, although not the exact recoil approximation, has been proposed in the \textsc{Hybrid} model~\cite{Casalderrey-Solana:2016jvj}. Instead of a perturbative description of parton scatterings with the QGP, this model modifies the energy-momentum of partons provided by the \textsc{Pythia} vacuum shower according to the holographic calculations of parton energy loss in a strongly coupled plasma~\cite{Chesler:2014jva}. The corresponding energy-momentum loss from each jet parton is then assumed to thermalize with the QGP instantaneously and is directly converted to hadrons using the Cooper-Frye formalism.

These different implementations correspond to different limits in the evolution of medium response. The \textsc{Jewel} model represents the non-interacting limit where recoil partons do not suffer additional scatterings with the medium, while the \textsc{Hybrid} model represents the opposite limit of extremely strong interactions in which energy-momentum transfer from jet to medium suddenly thermalizes and hadronizes. The \textsc{Lbt} model is in between. It allows perturbative scatterings of the recoil partons with the medium before they hadronize at the QGP phase boundary. Therefore, while they provide qualitatively consistent conclusions on effects of medium response on jet observables, quantitative differences can be expected in the strength of these effects.

Besides the linear Boltzmann approach that implements parton transport inside a hydrodynamic medium, full Boltzmann transport models have also been developed for simulating the evolution of jet partons and medium partons within the same perturbative framework. Although dispute remains on the applicability of weakly coupled approaches to the QGP evolution, it is reasonable for interactions between jet and medium partons considering the large energy scale of the former. These interactions transfer energy-momentum from jet to thermal partons, thus naturally including effects of medium response to jet energy loss. For instance, the \textsc{Ampt} model has been applied to study the effects of medium response on the dihadron and $\gamma$-hadron correlations~\cite{Ma:2010dv}, the transport of the lost energy from hard jets to the lower $p_\mathrm{T}$ hadrons~\cite{Gao:2016ldo}, and the enhancement of the baryon-to-meson ratio at intermediate $p_\mathrm{T}$ around the quenched jets~\cite{Luo:2021voy}. Similarly, the \textsc{Bamps} model has been used to explore the viscous effect on the Mach cone structure~\cite{Bouras:2014rea}.

\subsection{Coupled parton transport and hydrodynamic evolution}
\label{subsec:CoLBT}

Whether the strongly coupled hydrodynamic approach or the weakly coupled transport approach provides a better description of medium response depends on the scale of the deposited energy from jets. Recoil partons at low energy scale are able to thermalize quickly, thus becoming a part of the hydrodynamic medium. Contrarily, those at large energy scale can still be approximated with quasi-particles and are expected to scatter with the QGP perturbatively before they approach thermal equilibrium.

To take into account the evolution of jet-induced medium excitation at different scales, various coupled parton transport and hydrodynamics approaches have been developed in recent years.
For instance, Ref.~\cite{Tachibana:2017syd} has formulated a coupled jet-fluid model to study the evolutions of full jet shower and the traversed medium with energy-momentum exchange between them.
In this coupled approach, the full jet shower evolution is described by a set of transport equations including both collisional and radiative processes, while the medium evolution is simulated via relativistic hydrodynamic equations with source terms accounting for the energy-momentum deposited from the jet shower.
With the inclusion of jet-induced medium excitation, the jet-fluid model can naturally explain the redistribution of energy around the quenched jets as measured by the CMS Collaboration~\cite{CMS:2016cvr}.

Recently, a concurrent coupled \textsc{Lbt} and hydrodynamics (\textsc{CoLbt-Hydro}) model has been developed in Refs.~\cite{Chen:2017zte}, in which jet partons at hard scales evolve within \textsc{Lbt} while nuclear matter at thermal scale evolves within \textsc{CLVisc}. To combine these two approaches into a consistent framework, the Boltzmann equation in \textsc{Lbt} is re-written in the Milne coordinates. At each proper time step $(\tau,\tau+\Delta \tau)$, \textsc{CLVisc} provides the medium information, based on which \textsc{Lbt} simulates elastic and inelastic scatterings of jet partons inside the QGP. Using the information of the final state partons from these scatterings, the source term for the energy-momentum deposition is constructed as
\begin{equation}
\label{eq:CoLBTsource}
J^\nu = \sum_i \pm_i\frac{\theta(p^0_\mathrm{cut}-p_i \cdot u)dp_i^\nu/d\tau}{\tau (2\pi)^{3/2}\sigma_r^2\sigma_{\eta_s}}\times \exp\left[-\frac{({\vec x}_\perp-{\vec x}_{\perp i})^2}{2\sigma_r^2}-\frac{(\eta_s-\eta_{si})^2}{2\sigma_{\eta_s}^2}\right],
\end{equation}
in which $p^0_\mathrm{cut}$ defines the separation scale between quasi-particles and the medium background. Jet and recoil partons inside \textsc{Lbt} with energies ($p_i\cdot u$) below this scale in the local rest frame of the medium are assumed to instantaneously thermalize with the QGP and constitute this source term with the $+_i$ sign. Since all ``negative" partons are originally sampled from the thermal background, their momenta are all subtracted from this source term as denoted by the $-_i$ sign. To the contrary, jet and recoil partons with energy above $p^0_\mathrm{cut}$ are kept in \textsc{Lbt} for further perturbative interactions with the medium. When constructing the source term, the energy-momentum deposition or depletion ($\pm_i dp_i^\nu/d\tau$) from jets is smeared in the coordinate space with a Gaussian function in Eq.~(\ref{eq:CoLBTsource}). The width parameters of the Gaussian are set as $\sigma_r=0.2$~fm and $\sigma_{\eta_s}=0.2$, in consistence with those for generating the initial energy density distribution of the medium by smearing primordial partons from \textsc{Ampt} simulations in \textsc{CLVisc}. This source term then enters \textsc{CLVisc} evolution of the QGP at the next proper time step. Iteration of this algorithm provides a simultaneous evolution of QGP, jets and their interactions.

\begin{marginnote}[]
\entry{Separation scale}{it has been verified in Ref.~\cite{Chen:2017zte} that varying $p^0_\mathrm{cut}$ between 1 and 4~GeV has little impact on the final jet spectra.}
\end{marginnote}

Using the \textsc{CoLbt-Hydro} simulation~\cite{Chen:2017zte}, one can clearly observe the Mach cone pattern of the wave fronts caused by the energy deposition from the jet, while a diffusion wake behind the jet due to energy depletion. This leads to an enhanced particle yield in the forward direction of jet propagation, while a suppression in the backward direction, as will be discussed in detail later. Moreover, it has been noted that the energy density deposited by jets can be as large as that of the unperturbed QGP background at early time when jets are sufficiently energetic. Therefore, instead of the linear response approximation, a full hydrodynamic calculation is necessary for a quantitative investigation of jet-induced medium excitation.

Concurrent simulation of jets and QGP has also been implemented in Ref.~\cite{Pablos:2022piv}, where the energy loss of mini-jets is calculated using a strongly coupled approach~\cite{Chesler:2014jva} and deposited into the \textsc{Music} hydrodynamic model~\cite{Schenke:2010nt}. Due to the abundance of mini-jets in heavy-ion collisions, their energy deposition into the medium is found to affect the collective flow coefficients of soft hadrons emitted from the QGP, therefore, ought to be taken into account for a precise extraction of $\eta_\mathrm{v}/s$ of the medium. Furthermore, although jet partons can hardly be affected by the energy deposited by themselves since they usually travel faster than the speed of sound, concurrent simulation becomes necessary when multiple jets travel through the medium disturbed by each other within the same event.

\subsection{Thermalization of energy-momentum deposition}
\label{subsec:thermalization}

Although instantaneous thermalization of jet energy deposition has been assumed in many studies, a sudden spatial smearing of the deposited energy can break causality. The thermalization process of this deposited energy is a challenging problem, which may help place additional constraints on the interaction strength between off-equilibrium quasi-particles and the thermal medium.

This thermalization process has also been explored within both weakly coupled and strongly coupled approaches. For the former, one may apply transport equations to study how jet partons approach thermal equilibrium through elastic and inelastic scatterings. This has been conducted in Ref.~\cite{Iancu:2015uja} for the first time, in which it has been found that as jet partons approach a low energy scale that is comparable to the medium temperature, elastic scatterings start to overwhelm the medium-induced splitting process for their further evolution and drive them towards thermal equilibrium.

The competition between elastic and inelastic processes can be inferred from the Langevin equation
\begin{equation}
dp^i/dt=-\eta_\mathrm{D}v^i+\xi^i, \quad \langle \xi^i(t)\xi^j(t')\rangle=\frac{\hat{q}}{2}\delta^{ij}\delta(t-t'),
\end{equation}
where $\vec{\xi}$ represents the thermal random force, $\eta_\mathrm{D}$ is the drag coefficient which can be related to the momentum space diffusion coefficient (or jet quenching parameter) $\hat{q}$ via detailed balance $\hat{q}=4T\eta_\mathrm{D}$. The thermalization time of a jet parton with energy $E$ can then be estimated by $t_\mathrm{th}=E/\eta_\mathrm{D}=(E/T)t_\mathrm{rel}$, in which $t_\mathrm{rel}=4T^2/\hat{q}$ is defined as the relaxation time. Here, $t_\mathrm{th}$ can be understood as the time for jet partons to approach the thermal scale, while $t_\mathrm{rel}$ gives the time to approach thermal equilibrium for a parton at the thermal scale. For $\hat{q}/T^3\sim 5$~\cite{JET:2013cls} and medium temperature $T\sim 0.3$~GeV, $t_\mathrm{rel}$ is on the order of 1~fm. Combining with the relations between the branching (or splitting) time of jet partons and their transverse momentum broadening $\hat{q} t_\mathrm{br}=k_\perp^2$ and an estimation of this branching time $t_\mathrm{br}=2E/k_\perp^2$, one obtains $t_\mathrm{br}=\sqrt{2E/\hat{q}}$, thus, $t_\mathrm{br}/t_\mathrm{th}= \sqrt{T/(2E)}\sqrt{\hat{q}/T^3}/2$. Therefore, for high energy partons, medium-induced splitting dominates ($t_\mathrm{br}\ll t_\mathrm{th}$), while for low energy partons, thermalization takes shorter time than forming new gluons. For $\hat{q}/T^3\sim 5$, these two times are comparable around $T\sim E$. For a gluon starting with an energy of $25 T$ in the $z$ direction, it is found in Ref.~\cite{Iancu:2015uja} that the parton distribution behind the leading parton by a distance of $t_\mathrm{rel}$ is already close to equilibrium around a time of $2t_\mathrm{rel}$.
The kinetic approach has also been implemented in Refs.~\cite{Schlichting:2020lef,Mehtar-Tani:2022zwf} to study jet energy loss and thermalization within a unified framework.

Thermalization of jet energy deposition has recently been combined with a strongly coupled medium in Ref.~\cite{Tachibana:2020mtb}. Before reaching thermal equilibrium, the energy deposition from jets first evolves according to the causal relativistic diffusion equation~\cite{Aziz:2004qu}
\begin{equation}
\label{eq:causal}
\left[\frac{\partial}{\partial t}+\tau_\mathrm{rel}\frac{\partial^2}{\partial t^2}-D_\mathrm{diff}\nabla^2\right]j_i^\nu(x)=0,
\end{equation}
in which the energy deposition from parton $i$ is initialized as $j_i^\nu=\pm_i p_i^\nu\delta^{3}(\vec{x}-\vec{x}_i^\mathrm{dep})$ and $\partial j_i^\nu/\partial t =0$ at the deposition time $t=t_i^\mathrm{dep}$. Same as the previous discussion in Sec.~\ref{subsec:CoLBT}, $p_i^\nu$ here denotes the four-momentum of parton $i$, which can be a positive particle (with the ``$+$" sign) above a certain energy threshold ($p_\mathrm{cut}^0$) in the local rest frame of the medium, or a negative particle (with ``$-$"). Meanwhile, $\tau_\mathrm{rel}$ is the relaxation time and $D_\mathrm{diff}$ is the spatial diffusion coefficient; they satisfy $v_\mathrm{sig}=(D_\mathrm{diff}/\tau_\mathrm{rel})^{1/2} < 1$ to respect causality. The solution of this $j_i^\nu$ at its thermalization time $t_\mathrm{th}$ contributes to the source term for the subsequent hydrodynamic evolution, i.e., $J^\nu (x)=\sum_i j^\nu_i(x) \delta(t-[t_i^\mathrm{dep}+t_\mathrm{th}])$. The model parameters above are set as $p_\mathrm{cut}^0=2$~GeV, $\tau_\mathrm{rel}=1.0$~fm, $D_\mathrm{diff}=0.6$~fm and $t_\mathrm{th}=1.5$~fm in Ref.~\cite{Tachibana:2020mtb}. Within this approach, effects of medium response on the angular distribution of jet energy have been compared between different implementations of medium response. It has been found that the difference between weakly coupled (recoil method in Sec.~\ref{subsec:recoil}) and strongly coupled (hydrodynamic response in Secs.~\ref{subsec:hydroResponse} and~\ref{subsec:CoLBT}) approaches depends on the evolution profile of the bulk medium. While their difference is moderate inside a static medium, sizable difference has been observed in the presence of the hydrodynamic flow.
This provides a first quantitative exploration of differences between various medium response models and suggests the importance of accurate dynamical modeling of the soft medium in studying jets.

\section{EFFECTS OF MEDIUM RESPONSE ON JET OBSERVABLES}

\subsection{Jet production}
\label{subsec:RAA}

The most frequently used observable for quantifying medium effects on jets is the nuclear modification factor defined as
\begin{equation}
\label{eq:RAA}
R_\mathrm{AA}(p_\mathrm{T},y,\phi)\equiv\frac{1}{\langle N_\mathrm{coll}\rangle}\frac{d^2N_\mathrm{AA}/(dp_\mathrm{T} dy d\phi)}{ d^2N_\mathrm{pp}/(dp_\mathrm{T} dy d\phi)},
\end{equation}
which gives the ratio of hadron or jet spectra between nucleus-nucleus (A+A) and proton-proton (p+p) collisions, with $\langle N_\mathrm{coll}\rangle$ being the average number of binary nucleon-nucleon collisions in each A+A collision. This $R_\mathrm{AA}$ has been compared between nearly all jet energy loss calculations and experimental data, from which one can extract the coupling strength between energetic partons and the QGP, or the jet transport coefficient $\hat{q}$~\cite{JET:2013cls,JETSCAPE:2021ehl,Xie:2022ght}.

While the $R_\mathrm{AA}$ of high $p_\mathrm{T}$ hadrons is mainly driven by the parton energy loss inside the QGP, the $R_\mathrm{AA}$ of fully clustered jets is much more complicated. Jet construction in heavy-ion collisions requires subtracting the QGP background. In reality, signals of jet partons always overlap with the medium background to some extent. Therefore, the clustered jets inevitably contain energy-momentum brought from the medium. As a result, the jet $R_\mathrm{AA}$ depends on both jet energy loss and the energy gain brought by recoil partons excited from the medium background. It has been confirmed in Refs.~\cite{Tachibana:2017syd,He:2018xjv,He:2022evt} that medium response has non-negligible impact on the jet $R_\mathrm{AA}$.

\begin{figure}[tbp]
    \centering
    \includegraphics[width=0.92\textwidth]{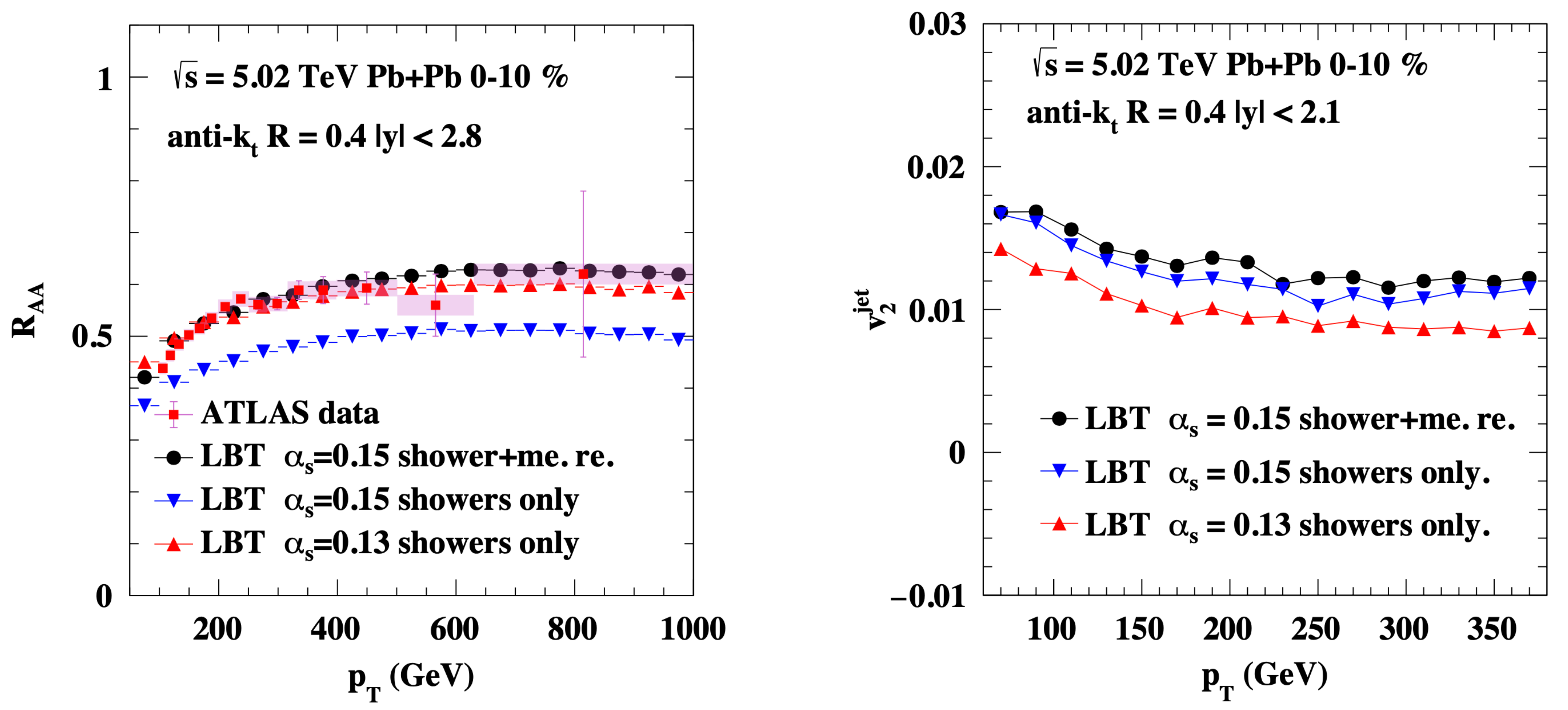}
    \caption{The jet $R_\mathrm{AA}$ (left) and $v_2$ (right) in Pb+Pb collisions at $\sqrt{s_\mathrm{NN}}=5.02$~TeV from \textsc{Lbt} calculations, compared between including and not including effects of medium response, and the ATLAS data~\cite{ATLAS:2018gwx}. The figures are from Ref.~\cite{He:2022evt}.}
    \label{fig:RAA-v2}
\end{figure}

In the left panel of Fig.~\ref{fig:RAA-v2}, one can observe the effect of medium response on the jet $R_\mathrm{AA}$ within the \textsc{Lbt} model. In the most central 10\% (denoted as 0-10\%) of Pb+Pb collisions at $\sqrt{s_\mathrm{NN}}=5.02$~TeV, a good agreement in the jet $R_\mathrm{AA}$ can be achieved by including medium response (recoil and ``negative" partons here) and setting $\alpha_\mathrm{s}=0.15$. Excluding medium response leads to an underestimation of the realistic jet energy, therefore a much smaller $R_\mathrm{AA}$ when the same $\alpha_\mathrm{s}=0.15$ is applied. If one re-fits the experimental data without medium response, a smaller value $\alpha_\mathrm{s}=0.13$ is required.

Besides affecting the extracted jet transport coefficient, medium response can also influence the cone size dependence of the jet $R_\mathrm{AA}$. Because recoil partons tend to be scattered to larger angles with respect to the incoming jet parton than medium-induced gluons, as one increases the cone for clustering jets, more energy-momentum from medium response is included in jets, causing an enhancement in their $R_\mathrm{AA}$. As found in Ref.~\cite{Tachibana:2017syd,He:2018xjv}, when medium response is not taken into account, small separation is seen between the $R_\mathrm{AA}$ of jets with different cone sizes from $R=0.2$ to 0.5. Contrarily, after including medium response, a significant increase is observed in the jet $R_\mathrm{AA}$ from smaller to larger cone sizes.
Similar cone size dependence has also been observed in other model calculations that incorporate medium response, as summarized in Ref.~\cite{CMS:2021vui}. Although the current CMS data at large $p_\mathrm{T}$~\cite{CMS:2021vui} seem to favor models without including medium response over those with medium response, discrepancy exists between the ATLAS~\cite{ATLAS:2018gwx} data and CMS data when the jet cone size is not very large, which needs to be resolved before  a firm conclusion can be drawn.

\subsection{Jet anisotropy}
\label{subsec:v2}

While the angular integrated $R_\mathrm{AA}$ quantifies the average $p_\mathrm{T}$ loss of jet partons or fully constructed jets, their collective flow coefficients $v_n$ characterize the azimuthal asymmetry of their transverse momentum distribution. The $v_n$ of high $p_\mathrm{T}$ hadrons is mainly determined by the asymmetric parton energy loss along different paths through a geometrically anisotropic QGP medium. At low $p_\mathrm{T}$, partons can quickly thermalize with the QGP and thus hadrons inherit the collective flow from the medium. In between, intermediate $p_\mathrm{T}$ hadrons are affected by the hadronization process that includes both fragmentation from higher $p_\mathrm{T}$ partons and coalescence between lower $p_\mathrm{T}$ partons~\cite{Zhao:2021vmu}. Unlike single inclusive hadrons, jets even at high $p_\mathrm{T}$ include both high and low energy constituent particles. As a result, their $v_n$ incorporates contributions from both parton energy loss and the QGP flow, and hence can be affected by jet-induced medium excitation.

The jet $v_n$ can be evaluated using either the event plane (EP) method or the scalar product (SP) method as
\begin{equation}
v_n^\mathrm{jet,EP}\equiv\langle\langle \cos[n(\phi^\mathrm{jet}-\Psi_n)]\rangle\rangle,\quad v_n^\mathrm{jet,SP}\equiv\frac{\langle\langle v_n^\mathrm{soft}\cos[n(\phi^\mathrm{jet}-\Psi_n)]\rangle\rangle}{\sqrt{\langle v_n^{\mathrm{soft}^2}\rangle}},
\label{eq:vn}
\end{equation}
in which the symbol $\langle\langle\ldots\rangle\rangle$ first averages jets within an event and then over different events, $\phi^\mathrm{jet}$ denotes the azimuthal angles of a jet, $\Psi_n$ is the $n^\mathrm{th}$-order event-plane determined by soft hadrons emitted from the QGP, and $v_n^\mathrm{soft}$ is the $v_n$ of soft hadrons. According to the definitions above, event-by-event fluctuations of $v_n^\mathrm{soft}$ can affect $v_n^\mathrm{jet,SP}$ but not $v_n^\mathrm{jet,EP}$. However, this effect is found small due to the small values of $\delta v_n^\mathrm{soft}/v_n^\mathrm{soft}$ from hydrodynamic calculations. These two methods give consistent $v_2$ and $v_3$ of both inclusive hadrons and jets within \textsc{Lbt}~\cite{Cao:2017umt,He:2022evt}.

As shown in the right panel of Fig.~\ref{fig:RAA-v2}, sizable effect of medium response has been found on the jet $v_2$~\cite{He:2022evt}. With a fixed value of $\alpha_\mathrm{s}=0.15$, the jet $v_2$ becomes smaller if medium response is excluded from jet construction, which results from two competing effects. On the one hand, medium response is stronger along longer path of jet propagation than shorter path. Therefore, removing medium response further increases the difference of the final jet $p_\mathrm{T}$ between longer and shorter paths, thus increasing the jet $v_2$. On the other hand, within jets, the collective flow carried by soft components close to thermal equilibrium with the QGP is larger than that of hard components caused by asymmetric energy loss. Therefore, excluding contributions from medium response reduces the jet $v_2$. Overall, the second effect dominates, especially when the medium flow is strong. Since $v_2$ is more complicated than $R_\mathrm{AA}$, it is conventional to study effects on the jet $v_2$ while keeping their $R_\mathrm{AA}$ fixed. As demonstrated in Fig.~\ref{fig:RAA-v2}, while the $R_\mathrm{AA}$ from \textsc{Lbt} excluding medium response with $\alpha_\mathrm{s}=0.13$ agrees with that including medium response with $\alpha_\mathrm{s}=0.15$, the former exhibits a $20\sim 40\%$ smaller $v_2$ than the latter. This difference increases as the jet cone becomes larger~\cite{He:2022evt}.

\subsection{Jet structures}
\label{subsec:substructure}

Advances in theoretical calculations and experimental measurements allow us to extend jet studies from the spectra of their total energy (or $p_\mathrm{T}$) to the energy distribution within jets. This is known as the jet structure or substructure.

The jet fragmentation function~\cite{CMS:2014jjt,ATLAS:2017nre} measures the fractional momentum distribution of particles within a jet, defined as
\begin{equation}
\label{eq:frag}
D(z)\equiv \frac{1}{N_\mathrm{jet}}\frac{dN_\mathrm{track}}{dz},
\end{equation}
where $z=\vec{p}^\mathrm{\;track}_\mathrm{T}\cdot\vec{p}^\mathrm{\;jet}_\mathrm{T}/|\vec{p}^\mathrm{\;jet}_\mathrm{T}|^2$ denotes the longitudinal momentum fraction of a particle with respect to the jet. To increase the resolution of particle distribution at low $p_\mathrm{T}$, a variable $\xi=\ln (1/z)$ is usually applied and $dN_\mathrm{track}/dz$ is replaced by $dN_\mathrm{track}/d\xi$ in Eq.~(\ref{eq:frag}). For $\gamma$/$Z$-triggered jets~\cite{CMS:2018mqn,ATLAS:2019dsv}, the momentum fraction $z$ can also be defined with respect to the $p_\mathrm{T}$ of $\gamma$/$Z$ as $z=\vec{p}^\mathrm{\;track}_\mathrm{T}\cdot\vec{p}^\mathrm{\;\gamma/Z}_\mathrm{T}/|\vec{p}^\mathrm{\;\gamma/Z}_\mathrm{T}|^2$, considering that they are not modified by the QGP and thus serve as a better reference of the initial jet energy.

As shown in Refs.~\cite{Casalderrey-Solana:2016jvj,KunnawalkamElayavalli:2017hxo,JETSCAPE:2018vyw,Chen:2017zte,Chen:2020tbl}, one can observe an enhancement of the jet fragmentation function at small $z$ (or large $\xi$) in A+A {\it vs.} p+p collisions due to the soft particle production from jet-medium interactions, among which the jet-induced medium excitation has a considerable contribution. At large $z$ (or smaller $\xi$), an enhancement can also be seen for single inclusive jets, because the energy loss of jets can increase the momentum fraction of their high $p_\mathrm{T}$ constituents. This enhancement at large $z$ does not appear in $\gamma$/$Z$-triggered jets when $\gamma$/$Z$ is used as the reference for defining $z$.
The jet fragmentation function here is similar to the distribution of the momentum fraction of hadrons relative to the triggered $\gamma/Z$, as will be discussed in detail in the next section.

\begin{figure}[tbp]
    \centering
    \includegraphics[width=0.95\textwidth]{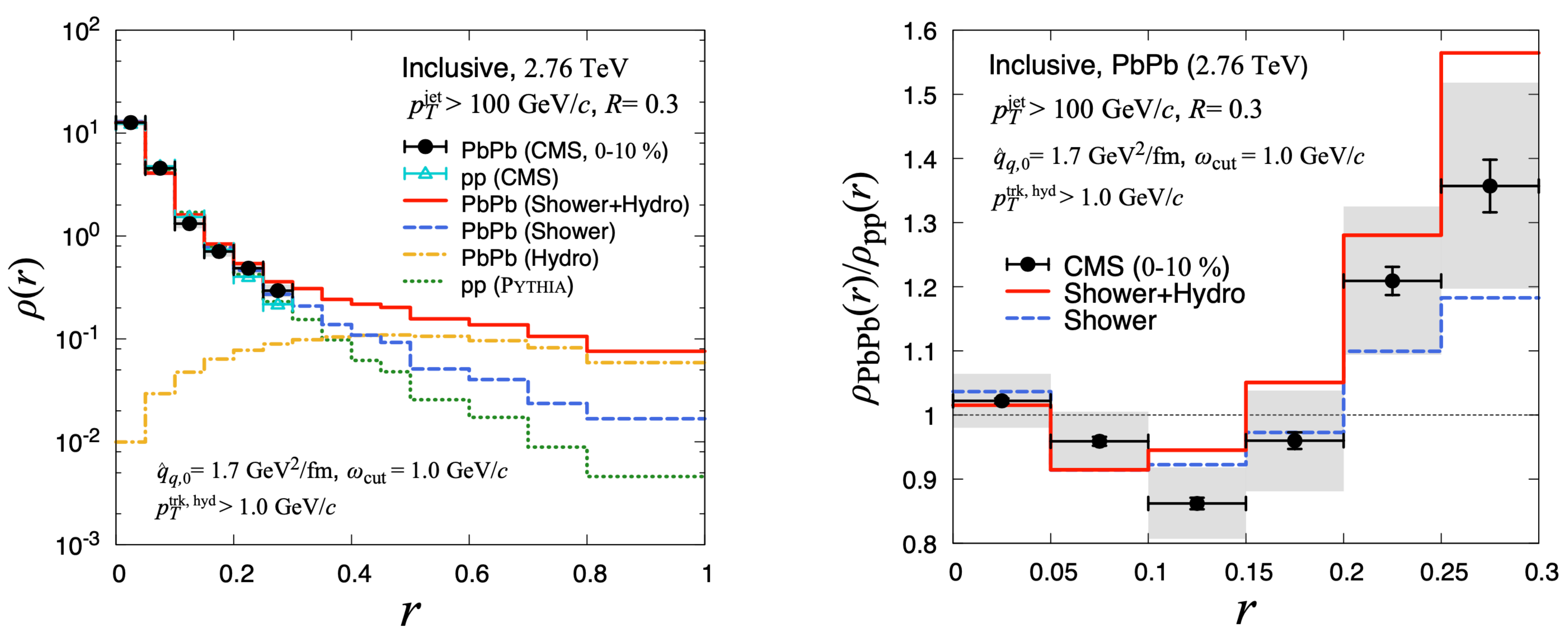}
    \caption{The jet shape (left) and its nuclear modification factor (right) in central Pb+Pb collisions at $\sqrt{s_\mathrm{NN}}=2.76$~TeV from the jet-fluid model calculation, compared between contributions from parton shower and hydrodynamic response, and the CMS data~\cite{CMS:2013lhm}. The figures are from Ref.~\cite{Tachibana:2017syd}.}
    \label{fig:shape}
\end{figure}

Apart from the momentum fraction distribution along the jet axis, we can also study the momentum distribution with respect to the opening angle relative to the jet axis. This is known as the jet shape~\cite{CMS:2013lhm,CMS:2016cvr}, defined by
\begin{equation}
\rho(r)\equiv \frac{1}{\Delta r}\frac{1}{N_\mathrm{jet}}\sum_\mathrm{jet}\frac{p^\mathrm{jet}_\mathrm{T}(r-\Delta r/2, r+\Delta r/2)}{p^\mathrm{jet}_\mathrm{T}(0,R)},
\label{eq:shape}
\end{equation}
in which $r=\sqrt{(\eta-\eta_\mathrm{jet})^2+(\phi-\phi_\mathrm{jet})^2}$ denotes the radius to the center of jet $(\eta_\mathrm{jet},\phi_\mathrm{jet})$ in the momentum space, and $p^\mathrm{jet}_\mathrm{T}(r_1,r_2)$ represents the summed $p_\mathrm{T}$ over particle tracks inside the circular annulus within ($r_1,r_2$). This jet shape is first normalized by the total $p_\mathrm{T}$ up to the cone size $R$ used for clustering the jet, and then normalized by the number of jets. Although jets are constructed using $R$, study of jet shape can be extended to $r>R$.

Shown in Fig.~\ref{fig:shape} is the jet shape in A+A and p+p collisions (left panel) and the corresponding nuclear modification factor (right panel) from a jet-fluid model calculation~\cite{Tachibana:2017syd}. In the left panel, contributions from different sources are compared for single inclusive jets in central Pb+Pb collisions at $\sqrt{s_\mathrm{NN}}=2.76$~TeV. At small $r$, the jet shape is dominated by energetic partons initiated from jets (labeled as ``shower"); however, at large $r$, it is dominated by hydrodynamic response to parton energy loss (labeled as ``hydro"). Compared to medium-induced gluon emission, elastic scattering together with hydrodynamic expansion has a wider angular distribution and thus is more effective in transporting jet energy to large $r$. In the right panel, one observes a significant enhancement of the nuclear modification factor at large $r$ after taking into account medium response. Similar findings have also been confirmed in Refs.~\cite{Casalderrey-Solana:2016jvj,Park:2018acg,JETSCAPE:2018vyw,KunnawalkamElayavalli:2017hxo,Luo:2018pto} for both inclusive jets and $\gamma$-triggered jets. For high $p_\mathrm{T}$ jets, small enhancement can also be observed in the jet shape at small $r$ due to the enhanced medium-induced splitting together with jet energy loss in A+A {\it vs.} p+p collisions. However, this enhancement at small $r$ will disappear for lower $p_\mathrm{T}$ jets~\cite{Chang:2016gjp,Chang:2019sae}.

Medium effect on the parton splitting function is the key to understanding the medium modification of parton showers. Direct measurements on the parton splitting function become available with the introduction of the soft drop jet grooming algorithm~\cite{Dasgupta:2013ihk,Frye:2016aiz}. In experiments~\cite{CMS:2017qlm,Kauder:2017cvz}, for a jet constructed using the anti-$k_\mathrm{T}$ algorithm with radius $R$, one may first re-cluster it using the Cambridge-Aachen algorithm and then de-cluster it in the reverse order by dropping the softer branch until two hard branches are identified with $z_g > z_\mathrm{cut}(\Delta R/R)^\beta$, where $z_g=\min(p_\mathrm{T1},p_\mathrm{T2})/(p_\mathrm{T1}+p_\mathrm{T2})$ is the fractional $p_\mathrm{T}$ of the softer branch. This jet grooming procedure suppresses contamination of soft hadrons from hadronization and underlying events in p+p collisions, thus connecting the perturbative calculation of the parton splitting function to the self-normalized $z_g$ distribution
\begin{equation}
p(z_g)\equiv \frac{1}{N_\mathrm{evt}}\frac{dN_\mathrm{evt}}{dz_g}.
\end{equation}

\begin{figure}[tbp]
    \centering
    \includegraphics[width=0.95\textwidth]{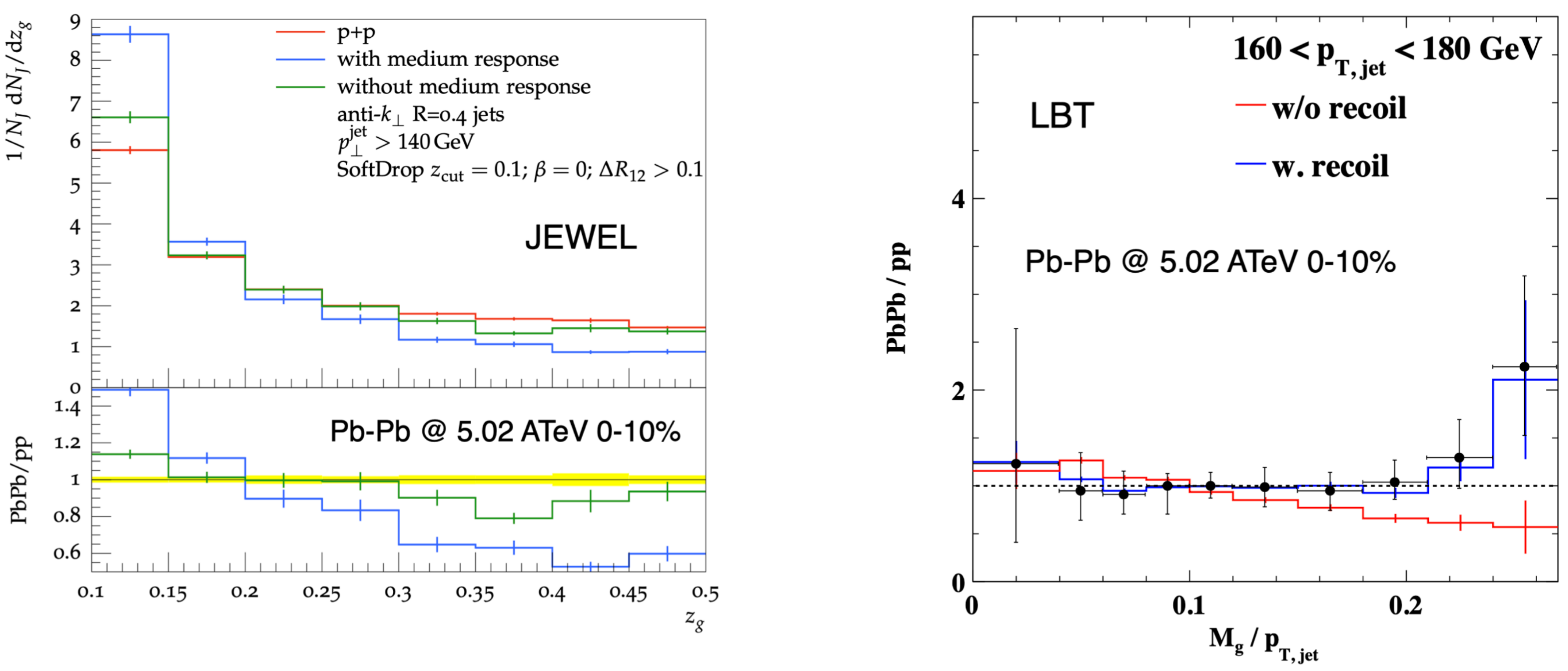}
    \caption{The $z_g$ distribution (left) and mass distribution (right) of groomed jets in central Pb+Pb collisions at  $\sqrt{s_\mathrm{NN}}=5.02$~TeV, compared between including and not including medium response, and the CMS data~\cite{CMS:2018fof}. The figures are from Refs.~\cite{Milhano:2017nzm,Luo:2019lgx}.}
    \label{fig:zg-mass}
\end{figure}

In A+A collisions, it has been shown in Refs.~\cite{Chien:2016led,Mehtar-Tani:2016aco,Chang:2017gkt,Li:2017wwc,Caucal:2019uvr} that a satisfactory description of this $p(z_g)$ can also be obtained within perturbative calculations on the medium-modified splitting function. However, it has also been suggested in Refs.~\cite{KunnawalkamElayavalli:2017hxo,Milhano:2017nzm} that soft hadrons contributed by jet-induced medium excitation can hardly be entirely removed using jet grooming and therefore still have an impact on $p(z_g)$ in A+A collisions. As shown in the left panel of Fig.~\ref{fig:zg-mass}, in central Pb+Pb collisions at $\sqrt{s_\mathrm{NN}}=5.02$~TeV, using parameters $z_\mathrm{cut}=0.1, \beta=0, \Delta R=0.1$ for jets constructed with $R=0.4$ and $p_\mathrm{T}$ greater than 140~GeV, one finds medium response significantly enhances $p(z_g)$ at small $z_g$ within the \textsc{Jewel} model. Since $p(z_g)$ is self-normalized, this enhancement leads to a faster decrease of its nuclear modification factor as $z_g$ increases. Contributions from medium response may be suppressed by adjusting the parameters of the jet grooming algorithm. Impacts from these parameters and different reclustering algorithms on $p(z_g)$ have been discussed in Ref.~\cite{Andrews:2018jcm}.

Medium effects on jets can also be reflected in the jet mass~\cite{ALICE:2017nij}, defined as
\begin{equation}
M=\sqrt{E^2-p_\mathrm{T}^2-p_z^2},
\label{eq:mass}
\end{equation}
with $E$, $p_\mathrm{T}$ and $p_z$ being the total energy, transverse momentum and longitudinal momentum of a given jet. As studied in Ref.~\cite{Park:2018acg,KunnawalkamElayavalli:2017hxo,Casalderrey-Solana:2019ubu}, the nuclear modification of jet mass comes from two competing effects. Jet energy loss in A+A collisions reduces their mass compared to jets with the same $p_\mathrm{T}$ in p+p collisions. On the other hand, jet-induced medium excitation introduces more soft components into jets and thus increases their mass. Overall, an enhancement of the average jet mass in A+A {\it vs.} p+p collisions can be seen after medium response is included. The enhancement becomes more prominent for jets with higher $p_\mathrm{T}$. Besides full jets, the mass of groomed jets has also been studied in Refs.~\cite{CMS:2018fof,Luo:2019lgx}. As shown in the right panel of Fig.~\ref{fig:zg-mass}, effects of jet-induced medium excitation could be large for jets with large masses. Without the contribution from recoil partons in \textsc{Lbt}, a suppression of the mass distribution is seen at large groomed jet mass due to jet energy loss. Contrarily, taking into account recoil partons leads to an enhancement.

Note that although we find medium response has noticeable effects on many jet observables, it is not the sole explanation of these observables. Even without medium response, large angle scattering and radiation~\cite{Lokhtin:2014vda,Perez-Ramos:2014fop,Chien:2015hda} and color coherence and de-coherence effects~\cite{Mehtar-Tani:2014yea} also provide reasonable descriptions of these observables. Therefore, it is necessary to further explore unique features of medium response, as will be discussed in the upcoming section.

\section{SIGNATURES OF MEDIUM RESPONSE IN JET-HADRON CORRELATIONS}
\label{sec:correlation}

While jet-induced medium excitation in relativistic heavy-ion collisions has been the subject of many recent studies, it is still a longstanding unresolved issue to search for the decisive signatures of medium response.
The identification, isolation and detailed study of medium response signals are of paramount significance for comprehensive understanding of jet-medium interactions and precise determination of the bulk properties of the strongly-interacting QGP.
Since the lost energy from jet shower partons is redistributed via rescatterings with medium partons, medium-induced radiations and jet-induced medium excitations, the medium response to jet energy loss should influence the production of soft and hard hadrons associated with the quenched jets.
Therefore, detailed analysis of hadron yields, spectra and distribution correlated with the triggered quenched jets should shed light on the signature of medium response.

\subsection{Jet-correlated hadron production}
\label{subsec:corrHadron}

In order to investigate the propagation and medium response concurrently in real time, Ref.~\cite{Chen:2020tbl} has developed the first coupled linear Boltzmann transport and hydrodynamics model \textsc{CoLbt-Hydro}.
In this approach, interactions of jet partons with the QGP is described by the \textsc{Lbt} model, while relativistic hydrodynamics is used to simulate the QGP evolution, with a source term describing the energy-momentum transferred from jets into the medium.

\begin{figure}[tbp]
	\centering
	\includegraphics[width=0.55\linewidth]{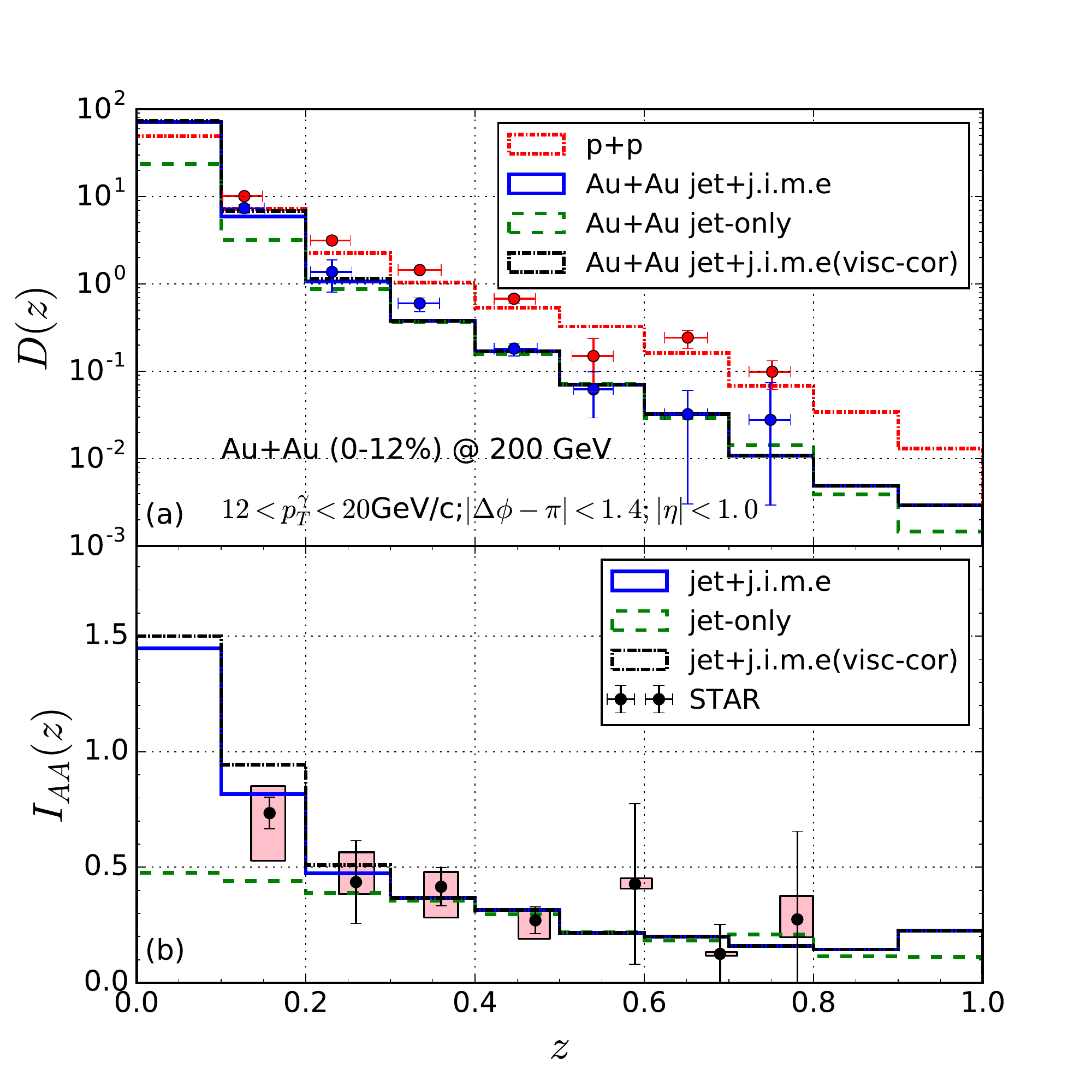}
	\caption{Upper: the $\gamma$-triggered jet fragmentation function $D(z)$ as a function of momentum fraction $z=p_\mathrm{T}^h/p_\mathrm{T}^\gamma$ for $p_\mathrm{T}^\gamma \in (12, 20)$~GeV, $|\eta|<1$ and $|\Delta \phi_{\gamma h} - \pi| < 1.4$, in p+p and 0-12\% Au+Au collisions at $\sqrt{s_\mathrm{NN}}=200$~GeV. Lower: The nuclear modification factor $I_\mathrm{AA}$ for the $\gamma$-triggered $D(z)$. The STAR data~\cite{STAR:2016jdz} are compared.
The figure is from Ref.~\cite{Chen:2017zte}.
}
	\label{fig:DAA-IAA}
\end{figure}

Based on the \textsc{CoLbt-Hydro} model, Ref.~\cite{Chen:2020tbl} studies the effect of jet-induced medium excitation on soft hadron production associated with the suppression of leading hadrons due to energy loss of hard partons in heavy-ion collisions.
Figure \ref{fig:DAA-IAA} shows the $\gamma$-triggered fragmentation function $D(z)$, with $z=p_\mathrm{T}^h/p_\mathrm{T}^\gamma$, in p+p collisions and central Au+Au collisions at $\sqrt{s_\mathrm{NN}}=200$~GeV (upper panel), together with their ratio $I_\mathrm{AA}=D_\mathrm{AA}/D_\mathrm{pp}$ (lower panel). The triggerred photon is set as $12< p_\mathrm{T}^\gamma < 20$~GeV, $|\eta| < 1$ and the associated hadrons are taken within $|\Delta \phi_{\gamma h} - \pi| < 1.4$. To compare with the STAR data, the zero-yield-at-minimum (ZYAM) method has been used to subtract a constant background. One sees that at intermediate and large $z$, \textsc{CoLbt-Hydro} can describe the STAR data on the suppression of leading hadrons due to parton energy loss. At small $z$ ($<0.1$), a significant enhancement is obtained for soft hadrons, contributed by jet-induced medium excitation.

\begin{figure}[tbp]
	\centering
	\includegraphics[width=0.9\linewidth]{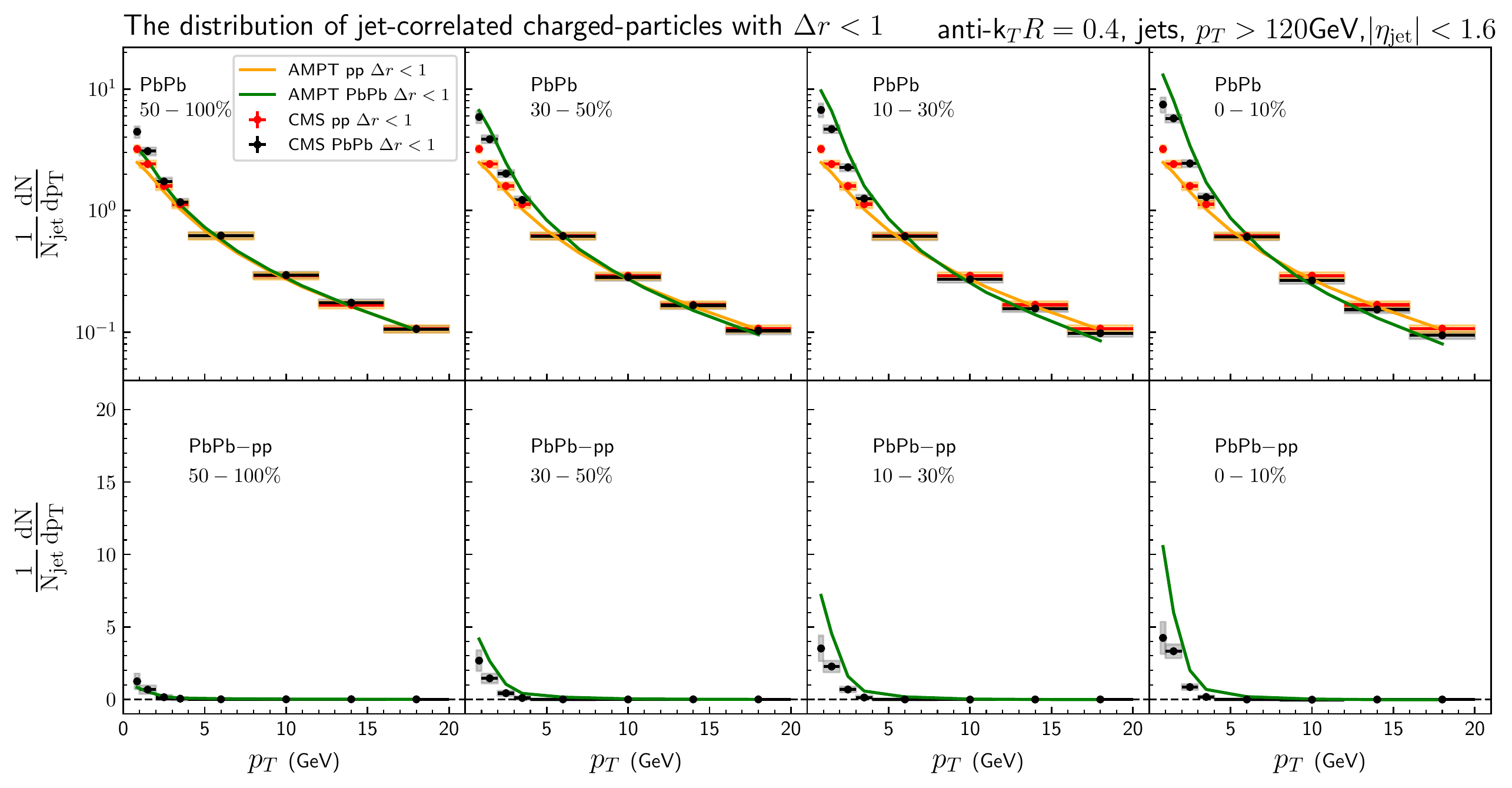}
	\caption{Upper: Jet-correlated charged particle spectrum within $\Delta r <1$ around triggered jets
in p+p and Pb+Pb collisions at $\sqrt{s_\mathrm{NN}} = 5.02$~TeV. Lower: The corresponding difference between Pb+Pb and p+p collisions.
The CMS data~\cite{CMS:2018zze} are compared.
The figure is from Ref.~\cite{Luo:2021hoo}.
}
	\label{fig:pT}
\end{figure}

In another work~\cite{Luo:2021hoo}, the production of jet-induced charged particles around the quenched jets is investigated using the \textsc{Ampt} model~\cite{Lin:2004en,Zhang:2005ni}.
The \textsc{Ampt} model provides a versatile tool to simulate the dynamical evolution of the bulk matter and jet-medium interactions for relativistic heavy-ion collisions.
In order to systematically subtract the background contribution and extract the jet-induced particle yield up to a very large distance from the jet direction, Ref.~\cite{Luo:2021hoo} uses the jet-particle correlation method following the CMS work~\cite{CMS:2016cvr,CMS:2021nhn}.
The mixed-event method is utilized to correct the limited acceptance effect for jet-particle correlations, and the side-band method is employed to subtract the background contribution from uncorrelated pairs and long-range correlations.

Figure \ref{fig:pT} shows the associated charged particle spectra $dN/dp_{\rm T}$ around the hard jets in p+p and Pb+Pb collisions at $\sqrt{s_{\rm NN}} = 5.02$~TeV.
The lower panel shows the difference between Pb+Pb and p+p events.
The triggerred jet is set as $p^{\rm jet}_\mathrm{T} > 120$ GeV, $R = 0.4$ and $| \eta_{\rm jet} | < 1.6$, and the associated particles are taken within $\Delta r = \sqrt{(\Delta \phi)^2 + (\Delta \eta)^2}<1$ with respect to the jet direction.
One can see that the \textsc{Ampt} calculation gives a reasonable description of the CMS data \cite{CMS:2018zze} on the jet-correlated charged particle spectra.
Compared to p+p collisions, the yield of associated high $p_{\rm T}$ particles around the triggered jets in Pb+Pb collisions is suppressed due to energy loss of leading partons of jets.
In contrast, the yield of the associated low $p_\mathrm{T}$ particles around the triggered jets in Pb+Pb collisions is enhanced compared to p+p collisions, which can be understood as the effect of medium response to parton energy loss.
This effect is more pronounced in more central collisions due to stronger jet quenching.
The above result from \textsc{Ampt} is very similar to that from \textsc{CoLbt-Hydro}.
So the general observation is that while parton energy loss leads to the suppression of hard particle production, jet-induced medium excitation enhances the soft particle production around quenched jets.

\subsection{Diffusion wake}
\label{subsec:wake}

Jet-induced medium excitation consists of two typical phenomena: the wave front and the diffusion wake.
The wave front is the transport of recoil partons which are excited by hard jets from the medium, while the diffusion wake is the depletion of energy behind hard jets and recoil partons.
As shown in the previous subsection, the wave front leads to an enhancement of soft hadrons around the axis of a quenched jet.
In this subsection, we will show that the diffusion wake can lead to a depletion of soft hadrons in the opposite direction of the propagating jet.
This has been extensively explored in Refs.~\cite{Chen:2020tbl, Yang:2021qtl, Yang:2022nei}.

Reference~\cite{Chen:2020tbl} studies the $\gamma$-triggered hadron yield as a function of the azimuthal angle $\Delta \phi_{\gamma h}$ in p+p collisions and central Au+Au collisions at $\sqrt{s_{\rm NN}} = 200$~GeV, without the ZYAM background subtraction.
First, the hadron yield at large $p_\mathrm{T}^h$ is found to be significantly suppressed due to parton energy loss.
More interesting finding is about the angular distribution of soft hadrons.
On the near side of the jet direction, i.e., $\Delta\phi_{\gamma h} \sim (\pi/2, \pi)$, one observes a significant enhancement of soft particle production.
In the mean time, a depletion of soft particle production is observed on the opposite side of the jet direction, i.e., around the triggered $\gamma$ direction ($\Delta\phi_{\gamma h} \sim 0$).
While the enhancement on the near side of the jet is due to the wave front excited by the hard jet as discussed in the previous subsection, the depletion in the $\gamma$ direction (i.e., the away-side of the jet) is due to the diffusion wake left behind the propagating hard jet inside the QGP.
This is a unique signal of jet-induced medium excitation which cannot be achieved using models without considering medium response.

\begin{figure}[tbp!]
	\includegraphics[width=0.85\linewidth]{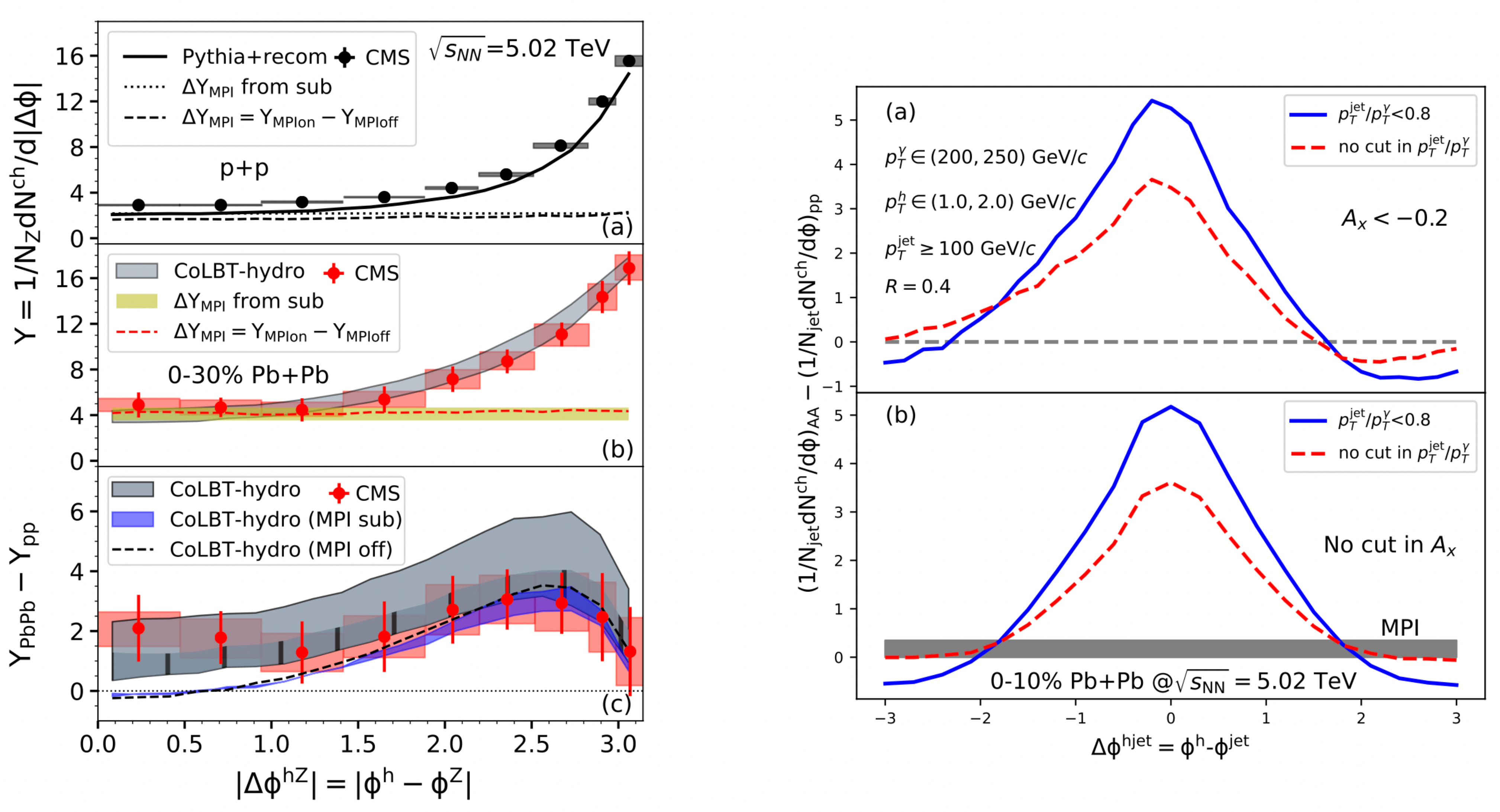}
	\caption{Left: The $Z$-hadron correlation as a function of the azimuthal angle for $p_\mathrm{T}^Z > 30$~GeV and $p_\mathrm{T}^h > 1~{\rm GeV}$ in p+p and 0-30\% Pb+Pb collisions at $\sqrt{s_{\rm NN}}=5.02$~TeV.
The CMS data~\cite{CMS:2021otx} are compared.
Right: The difference in $\gamma$-hadron correlation as a function of $\Delta \phi^{h\rm -jet}$ between 0-10\% Pb+Pb and p+p collisions.
The figures are from Ref. \cite{Yang:2021qtl}.
}
	\label{fig:Zh}
\end{figure}

Although it was predicted in Ref. \cite{Chen:2020tbl} that the diffusion wake can lead to a depletion of soft hadrons on the opposite side of the jet direction \cite{Luo:2021hoo}, recent CMS data on $Z$-hadron correlations showed an enhancement of soft hadrons in both $Z$ and jet directions~\cite{CMS:2021otx} instead. In order to decipher this intriguing result, Ref. \cite{Yang:2021qtl} has performed a detailed analysis on $Z$-hadron correlations within the framework of \textsc{CoLbt-Hydro} by considering the multi-parton interaction (MPI) effect in the initial nuclear collisions.

Figure \ref{fig:Zh} (left) shows the $Z$ boson triggered charged hadron yield as a function of the azimuthal angle $\Delta \phi^{hZ}$ in p+p and 0-30\% Pb+Pb collisions at $\sqrt{s_{\rm NN}} = 5.02$~TeV, together with their difference, compared to the CMS data~\cite{CMS:2021otx}.
The triggerred $Z$ boson satisfies $p_\mathrm{T}^Z > 30$~GeV and the associated hadrons satisfy $p_\mathrm{T}^h > 1$~GeV.
One can clearly see a peak in the jet direction ($\Delta \phi^{hZ} \sim \pi$), which is enhanced and broadened in Pb+Pb collisions relative to p+p collisions.
In the $Z$ direction ($\Delta \phi^{hZ} \sim 0$), we observe a sizable hadron yield in p+p collisions; it is also enhanced in Pb+Pb collisions.
This is a very puzzling result because jet-induced diffusion wake is expected to deplete hadron production in the opposite direction of jet propagation.
More detailed analysis by Ref. \cite{Yang:2021qtl} reveals that hadrons in the $Z$ direction mainly come from the MPI effect, such as independent production of mini-jets, associated with triggered hard processes, which typically give rises to a uniform distribution in the azimuthal angle for $Z$-hadron correlation.
Due to the interaction of these mini-jets with the medium, there is an enhancement of soft hadrons together with a suppression of hard hadrons from the MPI effect.
After subtracting the contribution from the MPI effect with a mixed event procedure, the signal of jet-induced diffusion wake becomes visible, as shown by a slight depletion on the near side of the $Z$-hadron correlation ($\Delta \phi^{hZ} \sim 0$).

\begin{figure}[tbp]
	\includegraphics[width=0.65\linewidth]{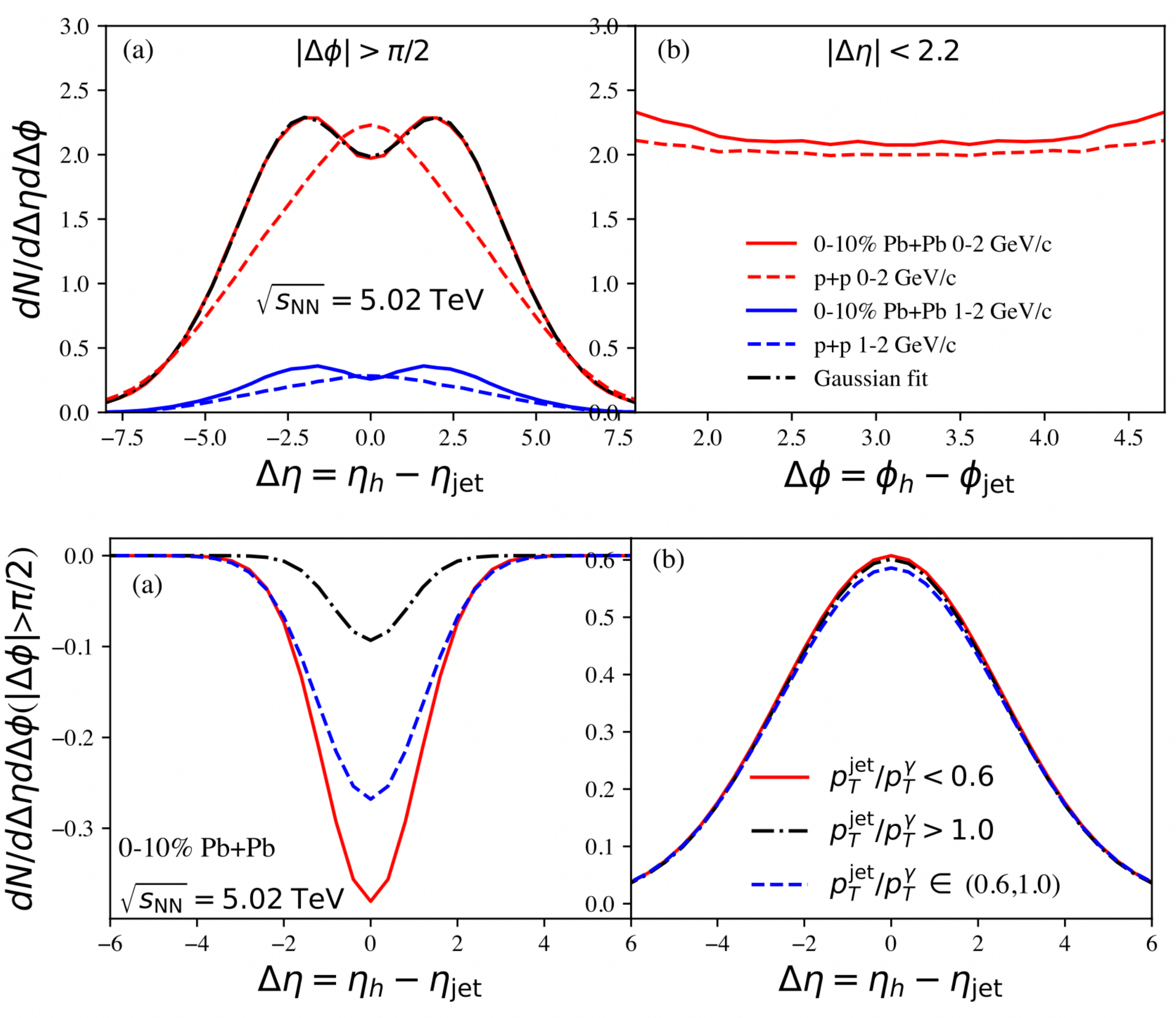}
	\caption{Upper: the $\gamma$-triggered jet-hadron correlation as a function of $\Delta \eta$ and $\Delta \phi$ in p+p and 0-10\% Pb+Pb collisions at $\sqrt{s_{\rm NN}}=5.02$~TeV. Lower: The diffusion wake valley and the MPI ridge in the $\gamma$-triggered jet-hadron correlation as a function of $\Delta \eta$ within $|\Delta \phi| > \pi/2$ for different ranges of $p_\mathrm{T}^{\rm jet}/p_\mathrm{T}^\gamma$ in Pb+Pb collisions.
The figures are from Ref.~\cite{Yang:2022nei}.
}
	\label{fig:valley-ridge}
\end{figure}

In order to have a more clear observation of the diffusion wake, Ref.~\cite{Yang:2021qtl} further proposes to use the longitudinal and transverse gradient jet tomography~\cite{Zhang:2007ja, Zhang:2009rn, He:2020iow} to localize the initial positions of the $Z/\gamma$-jet events.
Figure~\ref{fig:Zh} (right) shows the nuclear modification of the associated the soft hadron yield as a function of $\Delta \phi^{h\rm -jet}$ in $\gamma$-jet events in 0-10\% Pb+Pb collisions at $\sqrt{s_{\rm NN}}=5.02$~TeV.
Here the triggered $\gamma$ is chosen to emit primarily in the $-\hat{y}$ direction.
In the upper panel, one selects $\gamma$-jet events with a transverse asymmetry cut $A_x < -0.2$ that biases the initial transverse positions of $\gamma$-jet events towards the $-\hat{x}$ direction.
For this biased case, the enhancement of soft hadrons at $\Delta \phi^{h \rm -jet}<0$ is more pronounced while the depletion of soft hadrons at $\Delta \phi^{h\rm -jet}>0$ region is deeper.
Such asymmetric feature is due to the fact that jet-induced medium excitation is stronger in denser region of the medium, and at the same time, the diffusion wake is stronger in the opposite direction (less dense region).
One can also select $\gamma$-jet events with small values of the transverse momentum asymmetry (e.g., $p_\mathrm{T}^{\rm jet}/p_\mathrm{T}^\gamma<0.8$) to bias towards longer jet path length.
This selection further enhances the signal of the diffusion wake, which can now be clearly seen even without subtracting the MPI background.

Furthermore, Ref.~\cite{Yang:2022nei} explored the three-dimensional structure of the diffusion wake induced by $\gamma$-triggered jets in Pb+Pb collisions using the \textsc{CoLbt-Hydro} model.
The result is presented in Fig.~\ref{fig:valley-ridge}. The upper panels show the $\gamma$-triggered jet-hadron correlation as a function of rapidity and azimuthal angle in p+p collisions and 0-10\% Pb+Pb collisions at $\sqrt{s_{\rm NN}}=5.02$~TeV.
It is very interesting to observe a double-peak structure in the rapidity distribution of soft hadrons in the opposite direction of jets.
More detailed analysis by Ref.~\cite{Yang:2022nei} reveals that this double-peak structure is a combined effect of a valley caused by the diffusion wake and a ridge from the MPI effect, as shown in the lower panels.
By selecting different values of the $\gamma$-jet asymmetry $p_\mathrm{T}^{\rm jet}/p_\mathrm{T}^\gamma$, it is found that the depth of the diffusion wake increases with the jet energy loss.
This is another signature of diffusion wake without subtracting the MPI background.

\subsection{Hadron chemistry around jets}
\label{subsec:chemistry}

As shown in previous discussions, jet-induced medium excitation may flow to very large angles with respect to the jet direction and change the energy redistribution around the quenched jets. In particular, it can enhance the production of soft hadrons at very large angle ($\Delta r \sim 1$) away from the jet direction. In this subsection, we explore the hadron chemistry around the quenched jets. Since the lost energy from jets is deposited into medium partons, the relative yields or chemical compositions of particles produced from jet-excited partons around the quenched full jets should be different from those produced from the vacuum jets. This interesting phenomenon has been explored in Ref.~\cite{Luo:2021voy,Chen:2021rrp,Sirimanna:2022zje}.

\begin{figure}[tbp]
	\includegraphics[width=0.95\linewidth]{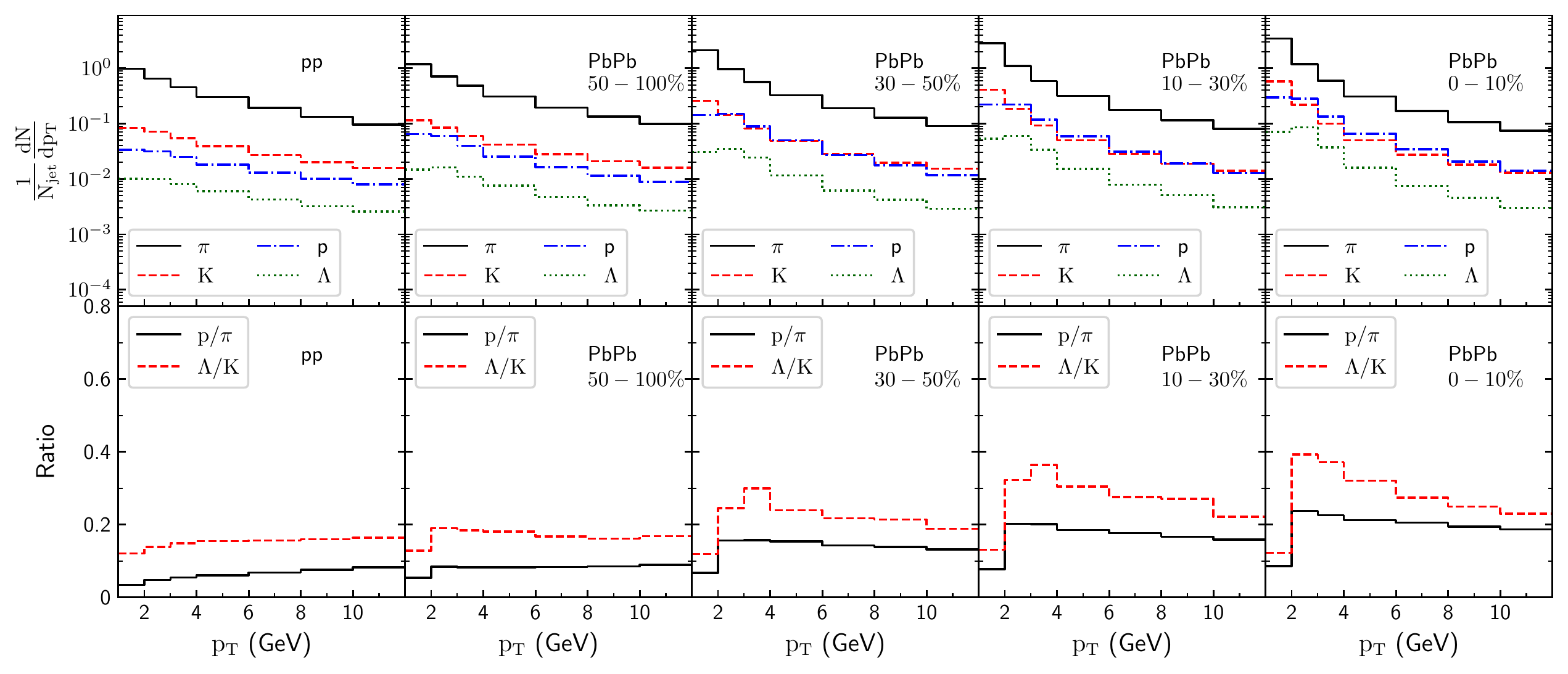}
	\caption{Upper: the jet-correlated $\pi$, $K$, $p$ and $\Lambda$ spectra around the triggered jets within $\Delta r <1$ in p+p and Pb+Pb collisions at $\sqrt{s_{\rm NN}} = 5.02$~TeV. Lower: The corresponding jet-correlated $p/\pi$ and $\Lambda/K$ ratios around the triggered jets.
The figure is from~\cite{Luo:2021voy}.
}
	\label{fig:dNdpT}
\end{figure}

Figure~\ref{fig:dNdpT} (upper) shows the jet-correlated $\pi$, $K$, $p$ and $\Lambda$ spectra ${dN}/{dp_\mathrm{T}}$ around the triggered jets within ${\Delta r = \sqrt{(\Delta \eta)^2 + (\Delta \phi)^2} <1}$ for p+p collisions and Pb+Pb collisions at $\sqrt{s_{\rm NN}} = 5.02$~TeV, calculated using the \textsc{Ampt} model.
The lower panel shows the corresponding jet-correlated $p/\pi$ and $\Lambda/K$ ratios around the triggered jets within $\Delta r <1$ as a function of the associated particle $p_\mathrm{T}$.
First, the yields of low $p_\mathrm{T}$ particles in Pb+Pb collisions are enhanced compared to p+p collisions.
Such enhancement is more pronounced in more central collisions due to the stronger jet quenching effect.
This is consistent with the previous finding that a significant amount of the lost energy from the quenched jets is carried by low $p_\mathrm{T}$ particles at large angles away from the jet direction~\cite{Tachibana:2017syd, Luo:2021hoo}.
Another extremely interesting observation is that the jet-correlated baryon-to-meson ratio in the intermediate $p_\mathrm{T}$ region around the triggered jets is strongly enhanced in Pb+Pb collisions relative to p+p collisions.
This is a very unique signature of medium response to jet quenching, and can be naturally explained by parton coalescence~\cite{Luo:2021voy}.
Due to interactions between hard jets and the QGP medium, the energy is deposited from jets to medium partons.
The coalescence between jet-excited medium partons tends to increase the relative yield of baryons to mesons at intermediate $p_\mathrm{T}$ around the quenched jets.

Since the lost energy can diffuse to very large angles with respect to the jet direction, the above jet-correlated baryon-to-meson enhancement at intermediate $p_\mathrm{T}$ around the quenched jets should have a strong dependence on the relative distance $\Delta r$, which has been explored in Ref.~\cite{Luo:2021voy}.
The nuclear enhancement (A+A minus p+p) of jet-correlated baryon-to-meson ratios is found to be stronger for larger $\Delta r$ (within $\Delta r < 1$).
This is because the lost energy from the quenched jets can flow to large angles, where the production of baryons relative to mesons is more enhanced due to parton coalescence.
Note that the parton coalescence mechanism, regardless of the details of various model implementations, has been very successful in explaining the number-of-constituent-quark scaling of elliptic flow and the enhanced baryon-to-meson ratios in the intermediate $p_\mathrm{T}$ region for the bulk matter in relativistic heavy-ion collisions~\cite{Fries:2003vb, Molnar:2003ff, Greco:2003xt, Hwa:2004ng}.
Therefore, the above qualitative prediction of the baryon-to-meson enhancement around the quenched jets should be robust. Experimental confirmation of this prediction will provide unambiguous evidence for medium response to jet quenching.

\section{CONCLUSIONS}
\label{sec:conclusion}

\begin{summary}[SUMMARY POINTS]
\begin{enumerate}
\item Jets serve as an energetic probe of the QGP created in relativistic heavy-ion collisions, and can reveal information on the transport properties and the underlying microscopic structures of the QCD matter.
\item Jet-medium interactions include both medium modification of jets and jet-induced medium excitation. Recent developments in theory and experiment allow us to extend jet studies from the former to the latter.
\item Both weakly coupled transport approach and strongly coupled hydrodynamic approach have been developed to study medium response to the energy-momentum deposition from jets and the thermalization process of this deposition.  Coupled transport and hydrodynamic models have been built for concurrent simulation of jet showers and QGP expansion, which is found to be successful in a simultaneous description of hard and soft components within jets.
\item Sizable effects of medium response have been observed on the nuclear modification factor and anisotropic flow coefficients of jets, as well as their fragmentation function, shape, splitting function and mass, indicating the essential role of medium response for precise extraction of the QGP properties from jet observables.
\item Jet-hadron correlations, including the diffusion wake on the opposite side of jet propagation and the enhanced baryon-to-meson ratio around the quenched jets, have been suggested as unique signatures of medium response, which can hardly be understood using theories without considering medium response.
\end{enumerate}
\end{summary}

\begin{issues}[FUTURE ISSUES]
\begin{enumerate}
\item A detailed comparison between various model implementations of medium response is necessary for a quantitative constraint on the systematic uncertainties in effects of medium response on jet observables.
\item Search for the diffusion wake and measurement on the hadron chemistry from the experimental side are urgent for a solid confirmation of the presence of jet-induced medium excitation in heavy-ion collisions.
\item Due to the fragile signal of medium response on top of the large and fluctuating QGP background, more precise calculations and measurements are desired for understanding the detailed mechanisms of medium response. This includes improving the hadronization model and the baseline of jet structures in p+p collisions in theoretical calculations, and resolving existing discrepancies in the jet data from different experimental measurements.
\item Developing schemes of enhancing signals of medium response, such as the longitudinal and transverse gradient jet tomography and application of machine learning techniques, will be helpful for a more direct extraction of the QGP properties, e.g. density, shear viscosity and speed of sound, from these signals.
\end{enumerate}
\end{issues}

\section*{DISCLOSURE STATEMENT}
The authors are not aware of any affiliations, memberships, funding, or financial holdings that
might be perceived as affecting the objectivity of this review.

\section*{ACKNOWLEDGMENTS}
We thank helpful discussions with Wei Chen, Xinrong Chen, Yichao Dang, Yayun He, Xiaowen Li and Yasuki Tachibana. This work was supported by the National Natural Science Foundation of China (NSFC) under Grant Nos.~12175122, 2021-867, 12225503, 11890710, 11890711, 11935007.

\bibliographystyle{ar-style5}
\bibliography{SCrefs}

\end{document}